\documentclass[prd,nofootinbib,twocolumn]{revtex4-2}
\date{\today}

\usepackage{graphicx}
\usepackage{bm}

\usepackage[english]{babel}

\usepackage{amsmath}
\usepackage{amsfonts}
\usepackage{amssymb}
\usepackage{tikz-cd}
\usepackage{mathrsfs}
\usepackage{xfrac}
\usepackage{url}
\usepackage{color}
\usepackage[colorlinks=true, citecolor=blue]{hyperref}
\usepackage{enumitem}
\usepackage[usestackEOL]{stackengine}
\usepackage{accents}

\newcommand{\bi}[3]{{#1}^{#2}_{\:\: {#3}}}

\newcommand{\si}[0]{\sin\!}
\newcommand{\co}[0]{\cos\!}
\newcommand{\ca}[1]{\mathcal{#1}}

\begin{document}

\title{General theory of swimming in curved spacetimes}
\author{Rodrigo Andrade e Silva}
\email{rasilva@umd.edu}
\affiliation{University of Maryland College Park,
College Park, MD 20742, USA}

\begin{abstract}
Swimming in curved spacetimes is a phenomenon whereby free bodies in curved spacetimes are able to propel themselves by performing cyclic internal motions. 
When originally proposed, it was further suggested that, in the limit of fast internal cycles, the net motion would display a simple geometric-phase character, in which the displacement per cycle would not depend on the time progression of the internal motions but only on the sequence of shapes assumed by the body, like a swimmer in a non-turbulent viscous fluid (low Reynolds number). 
In this paper we develop a general, covariant theory of swimming in curved spacetimes, describing a technique to study the motion of free, small, light, articulated bodies in general relativity by mapping the problem to an analogue in special relativity.
We give considerable attention to the limit of fast cycles and investigate the conditions in which the overall motion could display such geometric-phase behavior. The conclusion, however, is that this simple behavior is only realized in very specific circumstances, depending on the structure of the body, characteristics of internal motions, initial conditions, and symmetries of the spacetime; whereas, in general, our formulas predict a more complicated dynamics. 
\end{abstract}


\maketitle

\section{Introduction}
\label{sec:intro}

Swimming in curved spacetimes is a phenomenon in which small bodies evolving freely in curved spacetimes are able to propel themselves by performing cyclic internal motions. 
When first introduced \cite{Wisdom}, it was suggested that in the limit of fast cycles a curious geometric-phase behavior would be typically exhibited. In that situation, the overall motion would depend only on the sequence of shapes assumed by the body and not on how fast they were performed, similarly to a swimmer in a fluid at low Reynolds numbers (e.g., a human in honey or a microbe in water) \cite{Wilczek}.
However, a rigorous general treatment of this problem, consistent with general relativity, was still missing for several years following this proposal. 
In \cite{Rodrigo} we developed a careful treatment of this problem, showing that the original analysis, based on an effective Lagrangian approach, was inappropriate --- in particular, it would predict swimming in maximally-symmetric spacetimes, such as de Sitter space, which is known to be impossible from general arguments \cite{DixonI}. The present paper is supplementary to \cite{Rodrigo}, describing the theory in full detail and generalizing it to a great extent. 

The problem of interest consists of describing the trajectory of a light, small, articulated body, performing a given sequence of internal motions, in a curved spacetime. The body is light so that it can be regarded as a test body in a fixed background spacetime, it is small in comparison to the length scale set by the spacetime curvature, and it is articulated in the sense that it can adjust its shape in a controlled manner (say, it has arms which can be deliberately stretched or waved). In order to solve this problem we employ the theory of dynamics developed by Dixon \cite{DixonI, DixonII, DixonIII}. Dixon's formalism has the valuable property of being automatically consistent with the fundamental conservation equation $\nabla_a T^{ab} = 0$ of general relativity, which is built into the core of the formalism. In addition, the equations of motion are given in a convenient multipole-expansion form which makes clear how the overall motion is affected by different orders in $l/r$, where $l$ is the length scale of the body and $r$ is the characteristic length scale of the gravitational field. While our focus is to investigate conditions in which a low-Reynolds-number swimming behavior would be realized, our formulas apply to more general situations within the scope of Dixon's formalism (up to octupole order approximations). For reference, other papers on the subject, also employing Dixon's formalism, include \cite{Harte,Harte2,harte2023local,mosna2022chaotic,vesely2019glide,machado2024free} --- these consider up to quadrupole order approximations, and focus on other aspects of the problem, such as what can be practically achieved by an extended body swimming in gravitational fields. 

This paper consists of five main sections, which we briefly summarize here. In Sec.~\ref{sec:Dixon} we briefly review Dixon's formalism, which underpins our theory. We present the general definitions of linear and angular momenta, as well as a natural and essentially unique notion of center of mass in general relativity (with respect to which we will measure the ``net displacement'' of the swimmer), and then state the equations of motion (in terms of the multipole moments of the body).

In Sec.~\ref{octupole} we deduce from Dixon's equations an approximate equation for the trajectory of the center of mass of a general small body, depending on the net gravitational force and torque acting on it. Basically, we shall find expressions for the velocity of the center of mass of the body, correct up to order $(l/r)^3$, by keeping only terms up to octupole order in the expressions for the force and torque. The reason for going up to this order is that in some relevant situations, such as a tripod in Schwarzschild spacetime, the swimming effect appears only at such order.

In Sec.~\ref{sec:multipole} we discuss a convenient manner to prescribe the quadrupole and octupole moments, based on a local correspondence to a flat spacetime. In particular, we explain how to ``turn off'' the curvature in the evaluation of these moments and how to treat the dynamics of the internal motions in this auxiliary flat space. We then prove that the errors in this approximation are consistent with the octopole order degree of accuracy that we desire. This prescription is a fundamental ingredient in our theory of swimming, being the way in which one describes the body and, in particular, how its articulated appendages are programmed to move.  

In Sec.~\ref{sec:Swimming}, considering the equations for the trajectory of the body, we are able to formally study the effective dynamics induced by fast cyclic internal motions.  
With the general formalism established, we proceed to analyze a regime suitable for  {\sl swimming at low Reynolds number}.
We introduce a fiber bundle called the {\sl state space} whose base space is the space of internal configurations and the fibers represent global dynamical variables (including the position of the center of mass); in this language, swimming (at low Reynolds number) can be characterized as a lift from a path on the base to a path on the bundle. This language is convenient for two reasons: first, it emphasizes that the internal motions are pre-specified (input data), while the dynamical parameters evolve through Dixon's equations (output data); and second, it makes clear that the net change of the dynamical variables (which depend only on the initial and final points of the lifted curve) depends only on the shape of the curve on the internal space (and not on its time parametrization).
Although such geometric-phase swimming behavior can be realized in principle, we find that it is limited to very specific circumstances, normally involving symmetries of the body, the internal motions and the spacetime to be in accord.
The intuition is as follows. Imagine a person falling in a vertical non-uniform Newtonian gravitational field that is slightly stronger in their vicinity than on themselves; and say that in one situation they fall with their arms held against their body, while in another situation they start with their arms closed, extend them, and close them again. In the second situation, there is an additional displacement compared with the first, since there is an additional force during the time in which their arms are extended (feeling the stronger gravity). It is clear that in Newtonian gravity this displacement goes to zero in the limit of a very fast cycle, since the total time exposed to the stronger gravitational field goes to zero. Notice that this happens because, in Newtonian gravity, it is only the mass that ``feels'' gravity, and the mass is independent of the rapidity of the motions. In general relativity, on the other hand, the entire stress-energy tensor interacts with gravity. In particular, linear momenta and tensions can also ``feel'' the gravity, and they both grow with the speed of the internal motions. Thus, it is reasonable that in the limit of fast cycles the net displacement could approach a non-zero limit. The swimming at low Reynolds number effect is realized only when the linear momenta contributions are dominant and the internal motions are non-relativistic, as in this case the momenta grows linearly with speed, exactly compensating the shorter cycle times. In general, however, tensions also contribute, which often leads to a non-converging limit for the net displacement in fast cycles. 
In those cases, while the overall motion does not exhibit the strict geometric-phase character, sometimes it can still be described in a relatively simple manner: we can introduce the {\sl extended state space}, given as a bundle whose base space is the tangent bundle of the internal configurations and fibers represent the dynamical variables, so the dynamics can again be expressed as a lift from curves on the base to curves on the bundle. 

In Sec.~\ref{sec:Conditions} we comment on some aspects of our approximations. We note that some terms that need to be negligible in order to justify the swimming at low Reynolds number could be problematic in some cases. Namely, they could affect the displacement formula at second order in the amplitude of the internal motions. Since the net displacement after a small cycle depends on the ``area'' of the cycle (in the internal configuration space), the corresponding formula could become unreliable. When the displacement formulas are accurate only up to first order in the internal amplitudes, we say that the motion displays {\sl quasi-swimming} behavior. Thus it is important to check, in any given situation, whether the potentially problematic terms are indeed suppressed. For a quite general model of the body, which includes many interesting examples (e.g., a bipod on a sphere or a tripod in Schwarzchild), we describe generic conditions on the spacetime that ensures swimming (or quasi-swimming) at low Reynolds number, regardless of any other details about the body. We should note that quite generally, even outside the scope of our approximations, extended bodies are still generally able to propel themselves by performing swimming-like internal motions, but the resulting motion does not have a universal behavior.

Following the discussion section (Sec.~\ref{sec:Discussion}), we also include three appendices. In App.~\ref{sec:bitensors} we review the basic definition of bitensors, which is used in Dixon's formalism. In App.~\ref{sec:bipod} we apply our theory to a concrete example of a bipod moving on a 2-sphere (modelled as a submanifold of a cosmological spacetime $\mathbb R \times S^3$), showing that this situation exhibits swimming at low Reynolds number and moreover our formulas predict the expected result. In App.~\ref{sec:FLRW} we consider another example, of a tripod in Friedmann-Lema\^itre-Robertson-Walker (or FLRW) spacetime, where swimming at low Reynolds number is also realized (as long as the spacetime is not maximally-symmetric).

We use Latin letters for {\sl abstract tensor indices} and Greek letters for {\sl tensor components} in a basis. The {\sl curvature tensor} is defined by $\left( \nabla_a \nabla_b - \nabla_b \nabla_a \right)\omega_c = - R_{abc}^{\;\;\;\;\,\, d} \,  \omega_d$, and the {\sl Ricci tensor} and {\sl scalar curvature} are respectively defined by $R_{ac} = - R_{abc}^{\;\;\;\;\,\, b}$ and $R = R_a^{\;\; a}$. The signature of the metric is $(+,-,-,-)$ and the system of units is such that $c=G=1$.

\section{Dixon's formalism}
\label{sec:Dixon}

In a series of papers published in the 70's \cite{DixonI, DixonII, DixonIII}, W. G. Dixon developed a formal theory to describe the dynamics of extended bodies in general relativity, in an explicitly covariant way. This section is dedicated to briefly review some of the main results of his theory, which will be relevant in our subsequent construction.

Let $T^{ab}$ be the stress-energy tensor of the body under consideration, assumed to be conserved (i.e., 
$\nabla_a T^{ab}=0$), and define the world-tube $W$ of the body as the support of $T^{ab}$. Then, under reasonable
assumptions about $T^{ab}$ 
and somewhat weak conditions on the ``strength'' of the gravitational field \cite{Beiglbock},\footnote{It is assumed that the spatial extension of the 
body is finite and such that the convex hull ${\cal B}$ of the intersection  of its world-tube $W$ with any space-like
hypersurface $\Sigma$ is contained in a normal neighborhood of any point of ${\cal B}$. This means, in particular, that there is only one 
geodesic in ${\cal B}$ connecting any two spatially-separated points of the body. Moreover,
 wherever $T^{ab}\neq 0$, the momentum density is assumed to be time-like and future-directed according to {\it all} observers; i.e., 
$T_{ab}n^an^b > 0$ and
$T_{ab}T^b_{\,\;c}n^an^c > 0$ for all time-like vectors $n^a$. Finally, these conditions are
supposed to hold also for the  momentum density ``propagated'' from any point $x$ to any point $x'$, both in ${\cal B}$,
where the ``propagation'' is performed along the geodesic connecting $x$ and $x'$, according 
to maps $K^{a'}_{\;\;b}$ and $H^{a'}_{\;\;b}$ defined in Appendix \ref{sec:bitensors}.} the following general results hold:
\begin{itemize}
\item[(i)] At any point $x \in W$, there is a  {\sl unique} time-like, future-pointing, unit vector $n^a=n^a(x)$ 
such that the {\sl linear momentum} $p^a = p^{a}(x,n)$ of the body (defined \cite{DixonI} with respect to $x$ and $n^a$) is entirely in the
direction of $n^a$. In essence, this means that there is a (unique) family of observers (those with four-velocity $u^a= n^a$) 
according to whom the total {\sl spatial} momentum of the body is zero. In addition, at each point $x \in W$, there is a natural definition for the {\sl angular momentum} $S^{kl} = S^{[kl]} = S^{kl}(x,n)$ of the body;
\item[(ii)] There is a {\sl unique}, covariantly defined, time-like curve $z(s)$, contained in the convex hull of $W$, that can be naturally
identified as the world-line of the {\sl center of mass} of the body (here $s$ is the proper time along this world-line). More specifically, it is defined as the set of points $z$ which satisfy $p_k(z, n(z)) S^{kl}(z, n(z)) = 0$. It is worth mentioning that the tangent vector to this curve at $z(s)$, $v^a(s)$, is not, in general, parallel to $u^a(s) := n^a(z(s))$. We refer to $u^a$ and $v^a$ respectively as the {\sl dynamical} and {\sl kinematical} velocities of the body;
\item[(iii)] At each point $z(s)$ on the world-line of the center of mass, there are also natural definitions for the {\sl (higher) multipole moments} $J^{k_1 \cdots k_n pqlr}(s)$, $n \ge 0$, of the body.\footnote{We are referring to these as  {\sl higher} multipole moments because the linear and angular momenta, $p^k$ and $S^{kl}$, can also be regarded as multipole moments -- respectively, the monopole and dipole moments.} The set of tensors $\{p^k,S^{kl}\} \cup \{J^{k_1 \cdots k_n pqlr}; n \ge 0\}$ depends, at each $s$, only on the restriction of $T^{ab}$ to $W \cap \Sigma(s)$, where $\Sigma(s)$ is the space-like surface generated by all geodesics starting at $z(s)$ orthogonal to $u^k(s)$. Conversely, if the tensors in this set are given for all $s$, then $T^{ab}$ can be completely reconstructed;
\item[(iv)] The conservation equation $\nabla_a T^{ab}=0$ can be translated into precisely two dynamical equations for the linear and angular momenta,
\begin{eqnarray}
\frac{\delta p^k}{ds} &=& F^k
\nonumber\\
\frac{\delta S^{kl}}{ds} &=& 2 p^{[k} v^{l]}+Q^{kl} \, ,
\label{emFQ}
\end{eqnarray}
where $\delta/ds := v^k\nabla_k$ is the covariant derivative operator along $z(s)$, and $F^k$ and $Q^{kl}$ are respectively called the {\sl gravitational force} and {\sl torque} acting on the body (and they can be expressed in terms of 
couplings between the Riemann curvature tensor or its derivatives and the multipole moments);
\item[(v)] The {\sl proper mass} of the body, $M$, is defined as the scalar in the expression $p^k = M u^k$. If $\vartheta_{klqr}$ is the volume element of the spacetime (associated with the metric), then we define the {\sl spin vector} of the body as $S_k = \frac{1}{2}\vartheta_{klqr}u^l S^{qr}$, and it satisfies
\begin{align}
&\frac{\delta S^{k}}{ds} = - u^k \dot u_l S^l + Q^k \, ,
\label{emQ2}
\end{align}
where ${\dot u}^k = \delta u^k/ds$ and the {\sl torque vector} $Q^k$ is defined from $Q^{kl}$ similarly to the spin vector. Then, the {\sl misalignment} between $u^k$ and $v^k$ is given by
\begin{align}
u^{[k}v^{l]} =  \frac{1}{2M^2}\vartheta^{klqr} \bar F_q S_r + \frac{1}{M} u^{[k}Q^{l]q}u_q  \,\, ,
\label{uvmis}
\end{align}
where $\bar F^k = (\bi{\delta}{k}{l} - u^k u_l)F^l$. Note that this provides an equation for $v^k$ in terms of the linear momentum, spin, force and torque.
\item[(vi)] For each (if any) symmetry of the spacetime with generator  $\xi^a$, there is a conserved dynamical quantity $\chi_\xi$ given by
\begin{align}
\chi_\xi = p^k \xi_k + \frac{1}{2} S^{kl} \nabla_k \xi_l \,\, ,
\label{chi}
\end{align}
which can be understood as a consequence of Noether's theorem.
\item[(vii)]For a spatially small test-body, one can keep only a few multipole terms in the expressions for the force and torque. Up to octupole order, they are
\begin{align}
F^k =& \frac{1}{2} v^{l} S^{qr} {R^k}_{lqr}  +  \frac{1}{6} J^{qrlw} \nabla^k R_{qrlw} + 
\nonumber\\
& + \frac{1}{18} J^{qrwtl} \nabla^k \left( \nabla_q R_{r w t l} - 2 \nabla_t R_{qrwl} \right) \, ,
\nonumber\\
Q^{kl} =& \frac{4}{3} J^{qrw[k} {R_{qrw}}^{l]} +
\nonumber\\
& + \frac{1}{3}J^{qrwt[k} \left( 3 \nabla_q {R_{rwt}}^{l]} +  2 \nabla_t {R_{qrw}}^{l]} \right) \, ,
\label{emSN3}
\end{align}
and we shall refer to the three terms in the expression for the force as, respectively, $F^k_{spin}$, $F^k_{quad}$ and $F^k_{octu}$; and to the two terms in the expression for the torque as, respectively, $Q^{kl}_{quad}$ and $Q^{kl}_{octu}$.
\end{itemize}

\section{Octupole-order equations of motion}
\label{octupole}

The problem of solving for the dynamical evolution of a finite body in general relativity can be formulated as an initial-value problem with Dixon's equations. Hence, we shall assume that the initial position of the center of mass of the body $z(s_0)$ is known, as well as the initial linear momentum $p^k(s_0)$, the initial spin $S^k(s_0)$, and the initial multipole moments $J(s_0)$ (the indices are omitted). The worldline of the center of mass $\ell$ is parameterized as $z(s)$, where $s$ is the proper time along $\ell$, so that the tangent vector $v^k$ has unit modulus. 

In a coordinate system $\{ x^\mu \}$ around $z(s_0)$, the trajectory $z^\mu(s)$ of the center of mass of the body satisfies
\begin{equation}
v^\mu(s) = \frac{dz^\mu}{ds} \, ,
\end{equation}
where $v^\mu$ can be obtained algebraically from (\ref{uvmis}), in terms of other dynamical quantities such as $p^\mu$, $S^\mu$ and the $J$'s. This equation can then be solved in conjunction with (\ref{emFQ}), as long as we know how to evolve the $J$'s. The most direct approach for determining the moments is to translate the internal laws governing the substances constituting the body into differential equations for the $J$'s. In this case, depending on the order of the system of differential equations, the values of the $J$'s and the appropriate set of their time derivatives $dJ/ds$, $d^2\! J/ds^2$, etc must be provided at $s=s_0$. We shall, nonetheless, adopt a different approach to specify the moments (see next section), significantly more appropriate for our purposes. In our description, we will find a way to evaluate the multipole moments as functions of the internal motions performed by the body. Thus, the internal motions (and the $J$'s) should be seen as part of the input of the problem, not the output. 

Our task now is to find an explicit formula for the kinematical velocity $v^k$ in terms of dynamical quantities like $u^k$, the spin and the force and torque. This formula does not need to be exact, but it must be accurate up to order $(l/r)^3$. The motivation for assuming this degree of accuracy comes from $(i)$ the fact that the swimming effect is expected to be observed at this order\footnote{The Newtonian analysis suggests that the swimming effect should be observed at order $(l/r)^2$. The analysis of the motion of a bipod moving freely on a spacetime with topology $\mathbb R \times S^3$ shows that the swimming propulsion can indeed be at this same order. The study of a tripod falling aligned with the radial direction of Schwarzschild spacetime, in a specific setup, yields that the propulsion caused by the internal motions takes place at order $(l/r)^3$ only.} and that $(ii)$ the equations are still sufficiently simple and manageable (for instance, we can only keep terms up to octupole order in the equations of motion).

Let us briefly study the magnitudes, in orders of $l/r$, of the various terms in the equations of motion. Typically, in a (normal) coordinate system centered at $z(s)$, aligned with $u^k(s)$, the components of the energy-momentum tensor $T^{\mu \nu}$ have comparable magnitudes. The components of the multipole moments, in the same coordinate system, are usually \footnote{In the low ``internal'' velocities regime, some of these components can be much smaller than implied, so we can think of these magnitudes as upper bounds in the subsequent analysis.} of the following orders of magnitude
\begin{align}
p^\mu \sim M , \,\, S^{\mu \nu} \sim Ml , \,\, J^{\mu \nu \lambda \sigma} \sim Ml^2 , \,\,  J^{\mu \nu \lambda \sigma \rho} \sim Ml^3 \,\, , \nonumber
\end{align}
where $M = \sqrt{p^k p_k}$ is the total mass of the body and $l$ is its characteristic (spatial) length scale (as measured over the hypersurfaces $\Sigma(s)$ orthogonal to $u^k$). If it is possible to write the components of the curvature tensor and its first derivative as 
\begin{align}
R_{\mu \nu \lambda \sigma} \sim \frac{k}{r^2} \, , \quad \nabla_\rho R_{\mu \nu \lambda \sigma} \sim \frac{k}{r^3} \, ,
\end{align}
where $k$ is some dimensionless factor (not necessarily constant) and $r$ is some parameter with dimension of length (to be interpreted as the local radius of curvature), then each term in (\ref{emSN3}) has the respective order of magnitude
\begin{align}
&F^{\mu} \sim \frac{kM}{r} {\left( \frac{l}{r} \right)}  +  \frac{kM}{r} {\left( \frac{l}{r} \right)\!}^2 + \frac{kM}{r} {\left( \frac{l}{r} \right)\!}^3 \,\, ,
\nonumber\\
&Q^{\mu\nu} \sim  kM {\left( \frac{l}{r} \right)\!}^2 + kM {\left( \frac{l}{r} \right)\!}^3 \,\, ,
\label{emSN3order}
\end{align}
where the order-to-order separation, in the parameter $l/r$, is evident.

In order to find the desired formula for $v^k$, we first write
\begin{equation}
u^{[k} v^{l]} = \epsilon^{kl} \, ,
\end{equation}
where $\epsilon^{kl}$ is an anti-symmetric tensor corresponding to the right-hand side of equation (\ref{uvmis}). We see that it has order
\begin{equation}
\epsilon^{\mu\nu} \sim k \left[ {\left( \frac{l}{r} \right)\!}^2 + {\left( \frac{l}{r} \right)\!}^3 + {\left( \frac{l}{r} \right)\!}^4 \right] +  k \left[ {\left( \frac{l}{r} \right)\!}^2 + {\left( \frac{l}{r} \right)\!}^3 \right]   \,\, ,
\end{equation}
where the first three terms come from the contraction of the force and spin (inside the square brackets we have, respectively, the spin-order, the quadrupole-order and the octupole-order force terms); and the last two terms come from the torque contracted with the dynamical velocity (inside the square brackets we have, respectively, the quadrupole-order and the octupole-order torque terms). Thus, the tensor $\epsilon^{\mu\nu}$ is of order $(l/r)^2$ or higher.

Contracting $u^{[k}v^{l]}$ with $u_k$ we get
\begin{equation}
v^k u_k = 1 -  2 \epsilon^{kl} \epsilon_{rl} u_k u^r + \ca O (\epsilon^4) \,\, ,
\end{equation}
where $\ca O (\epsilon^4)$ represent terms that contain at least four factors of $\epsilon^{kl}$. Since we are only keeping terms up to order $(l/r)^3$, we can even neglect terms with two factors of $\epsilon^{kl}$, i.e., $\ca O (\epsilon^2)$, obtaining
\begin{equation}\label{vapprox2}
v^k \approx u^k + 2 \epsilon^{lk} u_l = u^k + \frac{1}{M^2} S^{kl} F_l + \frac{1}{M} Q^{kl} u_l  \, ,
\end{equation}
which is correct up to order $(l/r)^3$. 
(It is interesting to note that if we evaluate $u^{[k}v^{l]}$ from this approximate formula for $v^l$, we obtain $\epsilon^{kl}$ exactly.)

If the body were point-like and free, its worldline would be a geodesic of the spacetime and thus satisfy the geodesic equation $v^r \nabla_r v^k = 0$. However, since the body is not point-like, its worldline $\ell$ should satisfy a different equation, namely
\begin{equation}
v^r \nabla_r v^k = v^r \nabla_r u^k + v^r \nabla_r \left[ \frac{1}{M^2} S^{kl} F_l + \frac{1}{M} Q^{kl} u_l \right] \,\, .
\end{equation}
From equation (\ref{emFQ}) we obtain
\begin{align}
v^l \nabla_l v^k &= \frac{1}{M} (F^k - F^l u_l u^k) - \frac{3}{M^3} F^r u_r S^{kl} F_l
\nonumber\\
&+ \frac{1}{M^2} Q^{kl} ( 2 F_l - 3 F_r u^r u_l) + \frac{1}{M^2} u^k Q^{lr} F_l u_r 
\nonumber\\
&+ \frac{1}{M^2} S^{kl} v^r \nabla_r F_l + \frac{1}{M} v^r \nabla_r Q^{kl} u_l \, ,
\label{worldeq2}
\end{align} 
which is the desired equation for the worldline of the body. Note that the terms involving two factors of force or a product of a force and a torque implicitly contain two factors of curvature, which contributes with the small factor of order $k^2/r^4$. Since we are dropping terms of this order, this equation can be simplified to
\begin{equation}\label{worldeq}
v^l \nabla_l v^k = \frac{1}{M} (F^k - F^l u_l u^k) + \frac{1}{M^2} S^{kl} v^r \nabla_r F_l + \frac{1}{M} v^r \nabla_r Q^{kl} u_l 
\end{equation} 
which contains a force term, a term with the spin and the time derivative of the force, and a term involving the time derivative of the torque. The order of magnitude of the last two terms, containing the time derivatives of the force and the torque, cannot be precisely evaluated yet as we still do not have a prescription to explicitly calculate the multipole moments. Therefore, before going further with the analysis of this equation, let us discuss a proposal to specify the multipole moments.
\\

\section{Prescription of the multipole moments}
\label{sec:multipole}

This section is devoted to introducing the core element of the covariant theory of swimming: the special manner to prescribe the multipole moments of the body. Considering equation (\ref{worldeq}) for the worldline of the center of mass of the body, we can see that all the information concerning its geometric shape and internal motions have to be fully encoded in the multipole moments.

One manner to prescribe the multipole moments, as noted by Dixon in his papers, is to translate the internal differential equations governing the substances compounding the body into equations relating the multipole moments. For instance, Dixon discusses a reasonable definition for a rigid body in general relativity, where the components of the moments are assumed to be constant in a tetrad rotating in a way determined by the total spin of the body. In the general case, one could use the correspondence between the stress-energy tensor and the set of multipole moments, mentioned in item $(iii)$ of section (\ref{sec:Dixon}), to translate the dynamical differential equations from one description to the other. Explicitly, the (higher) multipole moments are given, for $n \ge 0$, by
\begin{equation}\label{explicJ2}
J^{k_1 \cdots k_n l p q r} = \Upsilon^{k_1 \cdots k_n [l [q p] r]} + \frac{1}{n+1} \Pi^{k_1 \cdots k_n [l [q p] r] t} v_t \, ,
\end{equation}
where $\Upsilon$ and $\Pi$ are respectively given, for $n \ge 2$, by
\begin{align}
&\Upsilon^{k_1 \cdots k_n lq}(s) = (-1)^n \! \int_{\Sigma(s)} \sigma^{k_1} \ldots \sigma^{k_n} \bi{\sigma}{l}{a} \bi{\sigma}{q}{b} T^{ab} w^c d\Sigma_c 
\nonumber\\
&\Pi^{k_1 \cdots k_n lqr}(s) = 2(-1)^n \! \int_{\Sigma(s)} \sigma^{k_1} \ldots \sigma^{k_n} {\langle n \rangle}^{prlq} H_{ap} T^{ab} d\Sigma_b
\label{UpPidef}
\end{align}
where $\Upsilon^{k_1 \cdots k_n lq}(s) = \Upsilon^{k_1 \cdots k_n lq}(z(s))$ and $\Pi^{k_1 \cdots k_n lqr}(s)$ are tensors at $z(s)$. The integral run over $x \in \Sigma(s)$, with $d\Sigma^c$ being the vector-valued element of volume induced on $\Sigma(s)$. The stress-energy tensor $T^{ab}$ is based at $x$, and $w^a$ is any vector field such that $w^a\delta s$ carries $\Sigma(s)$ to $\Sigma(s+\delta s)$. The quantities $\sigma^{k}$,  $\bi{\sigma}{l}{a}$,  $\bi{\sigma}{q}{b}$, $H_{ap}$ and ${\langle n \rangle}^{prlq}$ are bitensors, based at $(x,z(s))$, are defined in App.~\ref{sec:bitensors}. Note that we use a similar index notation convention for {\sl bitensors} as for tensors, with the difference that for bitensors based at $(x,z)$ the first half of the alphabet (``$a,b,c,\ldots$'' or ``$\alpha, \beta,\gamma,\ldots$'') is reserved for indices at $x$ (first entry) and the second half of the alphabet (``$k,l,m,\ldots$'' or ``$\kappa,\lambda,\mu,\ldots$'') for indices at $z$ (second entry).

So, how exactly one should proceed to specify these moments in a particular situation? In the case of the swimming bipod, for example, we could consider that a certain physical device is responsible for producing the desired internal movements. A simple realization is to consider a set of springs interconnecting the legs, so as to make them open or close as the springs oscillate; also, the legs could be considered to be springs themselves, so that they would be stretching and shrinking in an oscillatory fashion. We would then have to provide a local, covariant description for these springs. However, even the simplest model (of {\sl locally Hookean} springs\footnote{By locally Hookean spring we mean a relativistic model for a spring that obeys the Hooke's law in the local instantaneous rest frame of any infinitesimal piece of it \cite{spring}.}) is described by highly involved differential equations. Moreover, an important drawback of this type of approach is that the internal motions are not easily controllable. That is, it would be a considerable effort to find which structural modifications (such as changing the position or stiffness of the springs) would have to be implemented, at each time, in order to make the body perform the desired sequence of internal motions.

Therefore, we shall look for an alternative approach to the prescription of the multipole moments, particularly one that allows us to directly provide the sequence of internal motions rather than the underlying mechanisms necessary to produce them. As proposed by Wisdom, we might assume that there is some ingeniously-designed device, continuously coupled along the whole extension of the body, that applies the exact amount of tension to each of its points such as to produce a formerly specified sequence of internal motions. One way to conceive this is to imagine that the problem is first solved theoretically to determine the required internal forces that should be applied to each point, then all the information is stored in the memory of microcomputers spread all over the body, which will tell the local mechanisms of the device how to act in order to apply the correct tensions once the body is actually released to swim. We shall not be too concerned with the engineering details of the swimming mechanisms, as we are only interested in the resulting dynamics of a body going through a sequence of shape changes.

The basis of the covariant theory of swimming is the realization that, {\sl if only $(l/r)^3$ order of precision is required in the equations of motion, then the quadrupole and octupole moments can be evaluated as if the neighboring spacetime were flat}. This follows from the fact that curvature only affects the quadrupole and octupole moments at order $(l/r)^2$. Thus, since the lowest-order contribution to the equations of motion from these moments (that would come from their evaluation in a flat background) are already at order $(l/r)^2$, then curvature would only produce terms of order $(l/r)^4$ or higher in the equations of motion. Basically, we will treat the equations of motions perturbatively in the curvature, so that we first study the internal motions as if the spacetime were flat, computing the multipole moments accordingly, and then insert them into the Dixon's equations of motion. It should be stressed that this procedure plays a role only in the translation from the description of the internal states of a body in terms of an energy-momentum tensor to the corresponding description in terms of the multipole moments.

Before going through the justification of this claim, let us consider the case where the spacetime is actually flat. In Minkowski spacetime the equations of motion reduce to
\begin{align}
&\frac{\delta p^k}{ds} = 0
\nonumber\\
&\frac{\delta S^{kl}}{ds} = 0 
\nonumber\\
&u^{[k}v^{l]} = 0 \, ,
\end{align}
which imply that the mass, the dynamical velocity and the spin are conserved. Also, since $v^k = u^k$, the worldline of the center of mass $\ell$ is a straight line (i.e., geodesic). The internal motions have only to satisfy the conservation equation
\begin{equation}
\partial_a T^{ab} = 0 \, ,
\end{equation}
where $\partial_a$ is the derivative operator associated with any inertial (Cartesian) coordinate system. 

In inertial (Cartesian) coordinates, the bitensors appearing in (\ref{UpPidef}) become
\begin{align}
&\sigma^\kappa = - (x^\kappa - z^\kappa) 
\nonumber\\
&\sigma^\alpha = - (z^\alpha - x^\alpha)
\label{flatbi1}
\end{align}
\begin{align}
&\bi{\sigma}{\alpha}{\beta} = \bi{\delta}{\alpha}{\beta}
\nonumber\\
&\bi{\sigma}{\alpha}{\lambda} =  - \bi{\delta}{\alpha}{\lambda}
\nonumber\\
&\bi{\sigma}{\kappa}{\lambda} = \bi{\delta}{\kappa}{\lambda}
\nonumber\\
&\bi{\sigma}{\kappa}{\beta} = - \bi{\delta}{\kappa}{\beta} 
\end{align}
\begin{align}
&\bi{H}{\alpha}{\kappa} = \bi{\delta}{\alpha}{\kappa} 
\nonumber\\
&\bi{K}{\alpha}{\kappa} = \bi{\delta}{\alpha}{\kappa} 
\label{flatbi2}
\end{align}
\begin{equation}
\langle n \rangle^{\mu\nu\kappa\lambda} = \eta^{\mu (\kappa} \eta^{\lambda) \nu} \quad \text{for $n \ge 2$} \, ,
\end{equation}
where $\delta$ is the Kronecker delta and $\eta$ is the flat (Minkowski) metric.

Since the worldline of the center of mass is straight, the spatial hypersurfaces $\Sigma(s)$ are all parallel to each other, so that we may simply take the $w$-vector as
\begin{equation}
w^\mu = u^\mu = v^\mu  \,\, .
\end{equation}
Consider the particular inertial frame that is at rest with respect to the center of mass. Further, let it be located at the spatial origin of the coordinates. Thus, the worldline of the center of mass $\ell$ is described by the vertical line
\begin{equation}
z^\mu(s) = (s, 0, 0, 0) \,\, ,
\end{equation}
where $s$ is the proper time along it. The momentum-energy tensor can be decomposed\footnote{In this section, the indices $i,j,k,l,m$ may be used to indicate spatial indices of tensor components, so that they assume values from $1$ to $3$. Hopefully, it will be sufficiently clear where they are being used with this meaning and where the standard convention (abstract indices) is being used.} as
\begin{equation}
T^{\mu\nu} = \left(
\begin{array}{cccc}
\rho & g^1 & g^2 & g^3 \\
g^1 & h^{11} & h^{12} & h^{13} \\
g^2 &  h^{12} & h^{22} & h^{23} \\
g^3 & h^{31} & h^{32}  & h^{33}
\end{array}
\right) \,\, ,
\end{equation}
where $\rho$ is the mass density of the matter distribution, $g^i = (g^1, g^2, g^3)$ the density of linear 3-momentum, and the 3-tensor $h^{ij}$ is the total stress 3-tensor. If $u^i$ is the 3-velocity of a piece of matter at a given point, then the total stress 3-tensor $h^{ij}$ is related with the stress 3-tensor $\tau^{ij}$ by $h^{ij} = \tau^{ij} + g^i u^j$, and $\tau^{ij}$ is such that, for an infinitesimal oriented surface of area $\delta A$ and a direction defined by the unit 3-vector $\hat n = (n^1, n^2, n^3)$, then the 3-vector $\tau^{ij}n_j \delta A$ yields the force that the matter on the negative side of the surface applies to the matter on the positive side of it.

Inserting these quantities in the definitions of $\Upsilon$ and $\Pi$, we get the following expressions for the quadrupole and octupole moments in flat spacetime
\begin{align}
&J^{\mu\nu\rho\sigma} = \int_\Sigma d\Sigma \, r^{[\mu} r^{[\rho} \left( T^{\nu] \sigma]} + v^{\nu]} T^{\sigma] 0} + v^{\sigma]} T^{\nu] 0} \right)
\nonumber\\
&J^{\lambda\mu\nu\rho\sigma} = \int_\Sigma d\Sigma \, r^{\lambda} r^{[\mu} r^{[\rho} \left( T^{\nu] \sigma]} + \frac{1}{2} v^{\nu]} T^{\sigma] 0} + \frac{1}{2} v^{\sigma]} T^{\nu] 0} \right) \,\, .
\label{flatJ45}
\end{align}
where $r^\mu = x^\mu - z^\mu (s) = (0, x^1, x^2, x^3)$, with $x \in \Sigma(s)$, are the ``position vectors'' across the spatial surfaces $\Sigma(s)$. 
More explicitly, in terms of $\rho$, $g^i$ and $h^{ij}$, the non-vanishing components of the quadrupole moments are
\begin{align}
&J^{0i0j} = \frac{3}{4} \int d^3\! x\,  x^i x^j \rho
\nonumber\\
&J^{0ijk} = \frac{1}{4} \int d^3\! x \left( 3 x^i x^k g^j - 2 x^i x^j g^k - x^j x^k g^i \right)
\nonumber\\
&J^{ijkl} = \int d^3\! x\,  x^{[i} x^{[k} h^{j]l]}  \, ,
\label{Jijkl}
\end{align}
and for the octupole,
\begin{align}
&J^{m0i0j} = \frac{1}{2} \int d^3\! x\, x^m x^i x^j \rho
\nonumber\\
&J^{m0ijk} = \frac{1}{8} \int d^3\! x \, x^m \left( 4 x^i x^k g^j - 3 x^i x^j g^k - x^j x^k g^i \right)
\nonumber\\
&J^{mijkl} = \int d^3\! x\, x^m x^{[i} x^{[k} h^{j]l]}  \, ,
\label{Jmijkl}
\end{align}
where the functions $\rho$, $g^i$ and $h^{ij}$ are all evaluated at the point $(s, x^1, x^2, x^3)$. This makes clear how the different components of the moments depend on the dynamical quantities associated with the matter distribution: when there are two time-indices (i.e., $\mu = 0$) among the last four indices, then the component depends on the mass distribution of the matter; when there is only one time-index among the last four, then the component depends on the linear 3-momentum distribution; and when there is no time-index among the last four, then the component depends on the total stress distribution (which includes the 3-velocity, linear 3-momentum and the stress 3-tensor distributions).

We will show that the exact expressions for the multipole moments, when curvature effects are included, differ from the flat spacetime expressions (\ref{flatJ45}) only at order $(l/r)^2$. First let us argue that, in the suitable sense, the geometry of a bounded object\footnote{An object here may be seen as a subset of the space defined by a set of intrinsic geometric rules.} of size (length) $l$ is affected by curvature only at order $R l^2 \times l \sim k (l/r)^2 \times l$, if $R l^2 \ll 1$. It is clear that we need to define precisely what it means to ``turn on the curvature'' in the neighborhood of the object. Although there is not a unique way to define this operation, the exponential map provides a very convenient manner to relate the curved with the flat space, as explained next. 

Assume that there is some point $p \in \ca V \subset \ca M$, where $\ca V$ is any open normal neighborhood in $\ca M$ which contains the convex envelope of the geometrical object. Since $\ca V$ is a normal neighborhood of $p$, we have that, for each point $x \in \ca V$, there is a unique vector $\xi$ in $T_p\ca M$ such that $\exp_p \xi = x$. In other words, there is a diffeomorphism of a neighborhood of $T_p \ca M$ (containing the $0$ vector) onto $\ca V$ given by the exponential map at $p$. Also, there is a natural coordinate system on $\ca V$ induced from a Cartesian coordinate system on $T_p \ca M$: if $\xi$ is decomposed in an orthonormal vector basis $\{ e_\mu \}$ at $p$ as $\xi = \xi^\mu e_\mu$, then the coordinates $\xi^\mu$ can be assigned to the point $x$, i.e., we define $x^\mu = \xi^\mu$ (called {\sl Riemann normal coordinates}). On top of being a geometrically natural coordinate system, it has the interesting property that the coefficients of the Taylor expansion of the metric $g_{\mu\nu}(x)$, with respect to $x^\mu$ about $0$, are determined by the curvature and its (covariant) derivatives. For instance, up to second order we have 
\begin{equation}
g_{\mu\nu}(x) = \delta_{\mu\nu} + \frac{1}{3} R_{\mu \rho \nu \sigma}(0) x^\rho x^\sigma + \ldots \, ,
\end{equation}
which suggests a very natural way to implement the ``turning on'' of the curvature. We simply define a one-parameter family of metrics $g_\lambda$ by
\begin{equation}
[g_\lambda]_{\mu\nu}(x) = \delta_{\mu\nu} + \frac{1}{3} \lambda R_{\mu \rho \nu \sigma}(0) x^\rho x^\sigma + \ldots \,\, ,
\end{equation}
so that $\lambda = 0$ corresponds to a flat metric and $\lambda = 1$ corresponds to the actual metric of the space.

Let us now try to convince ourselves that curvature affects the basic laws of flat geometry at order $R l^2$ only (note that this is not intended to be a rigorous argument, in part because the proposition itself lacks the necessary precision). We shall consider a few basic geometrical constructions and try to identify the effect of curvature. For instance, note that the trajectory of a geodesic starting at $x_0^\mu$ with (unit) tangent vector $v^\mu$ is described by
\begin{equation}
x^\mu(s) = x_0^\mu + v^\mu s - \frac{1}{3} s^2 {R_{\nu \rho\sigma}}^\mu x_0^\nu v^\rho v^\sigma + \ldots \,\, ,
\end{equation}
where $s$ is the proper length along it and the curvature tensor is evaluated at the origin, $p$. If a geodesic defined by the same prescription lived in a flat space, then its trajectory $\widetilde x^\mu(s)$ would be given by
\begin{equation}
\widetilde x^\mu(s) = x_0^\mu + v^\mu s \,\, ,
\end{equation}
which indicates that they deviate from each other by
\begin{equation}
\Delta x^\mu (s) = x^\mu(s) - \widetilde x^\mu (s) = - \frac{1}{3} s^2 {R_{\nu \rho\sigma}}^\mu x_0^\nu v^\rho v^\sigma + \ldots \,\, ,
\end{equation}
We then see that the effect of the curvature $R \sim k/r^2$, in a geodesic of length $l$ within $\ca V$ (so that $ x_0^\mu \lesssim l$), is of order $l \times k(l/r)^2$, supporting the claim.
\begin{figure}[!h]
\centering
\includegraphics[scale = 0.4]{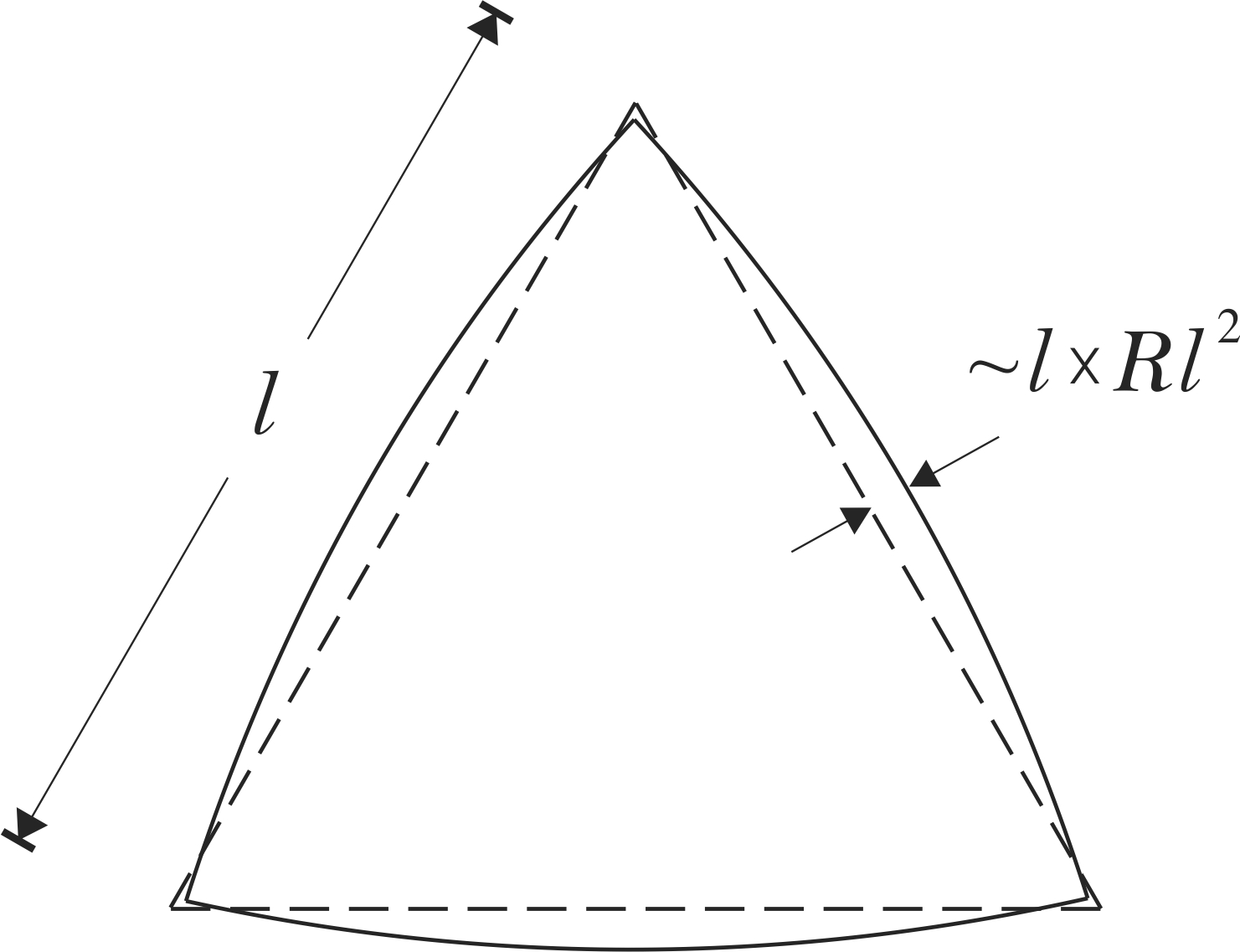}
\caption{Pictorial representation of an equilateral triangle being deformed as the curvature is turned on $-$ the dotted-line triangle corresponds to the flat-space object and the full-line triangle corresponds to the curved-space object. The triangle has edges of length $l$ and the deformation takes place at the scale $l \times Rl^2$.}
\label{FIGswimtriangle}
\end{figure}

For a few more examples \cite{Brewin,Brewin2}, consider the $(i)$ geodesic distance $L(x_1, x_2)$ between two points $x_1^\mu$ and $x_2^\mu$ in $\ca V$,
\begin{equation}\label{Lnormal}
L(x_1, x_2)^2 = \delta_{\mu\nu} \Delta x_{12}^\mu \Delta x_{12}^\nu + \frac{1}{3} R_{\mu\rho\nu\sigma} x_1^\rho x_1^\sigma x_2^\mu x_2^\nu + \ldots \,\, ,
\end{equation}
where $\Delta x_{ij}^\mu = x_j^\mu - x_i^\mu$; the $(ii)$ cosine law for a geodesic triangle with vertices $x_0$, $x_1$ and $x_2$,
\begin{align}
2 L_{01} L_{02} \co\theta_0 &= L_{01}^2 + L_{02}^2 - L_{12}^2 
\nonumber\\
&+ \frac{1}{3} R_{\mu\rho\nu\sigma} \Delta x_{01}^\rho \Delta x_{01}^\sigma \Delta x_{02}^\mu \Delta x_{02}^\nu + \ldots \, ,
\end{align}
where $\theta_0$ is the angle subtended at the vertex $x_0$ and $L_{ij} = L(x_i, x_j)$; and the $(iii)$ parallel transport of a vector $v_0^\mu$ at $x_0$, along an arbitrary curve $x^\mu(s)$ starting at $x_0$,
\begin{equation}
v^\mu(s) = v_0^\mu + \frac{1}{3} \left( {R_{\nu \rho\sigma}}^\mu + {R_{\nu \sigma\rho}}^\mu \right) v_0^\rho a_1^\sigma \left( s a_0^\nu + \frac{1}{2} s^2 a_1^\nu \right) + \ldots 
\end{equation}
where $a_0^\mu = x_0^\mu$ and $a_1^\mu = \left. dx^\mu/ds \right|_{s=0}$. This suggests that the geometry of the body can be well-represented in a flat space, given our approximations. 


Let us now examine other aspects of this approximation. Observe that the fundamental conservation equation, $\nabla_a T^{ab} = 0$, is only affected by the curvature at an order assumed to be negligible in Dixon's equations
\begin{align}
0 &= \nabla_\mu T^{\mu\nu}(x) = \partial_\mu T^{\mu\nu} + \Gamma^\mu_{\,\, \mu \sigma} T^{\sigma \nu} + \Gamma^\nu_{\,\, \mu \sigma} T^{\mu \sigma} \approx
\nonumber\\
&\approx \partial_\mu T^{\mu\nu} - \frac{1}{3} R_{\rho \sigma} x^\rho T^{\sigma \nu} + \frac{2}{3} {R_{\rho\mu\sigma}}^\nu x^\rho  T^{\mu \sigma} \, ,
\end{align}
which implies that $\partial_\mu T^{\mu\nu} \approx 0$ as the terms containing the curvature can be dropped (for they contain a factor of order $k/r^2$). Hence, the dynamics of the internal motions of the body can be described by special relativity.

The first curvature-related corrections to the flat spacetime expressions for the bitensors in (\ref{UpPidef}), if we particularize to $z = p$, are
\begin{align}
&\sigma^\alpha = x^\alpha 
\nonumber\\
&\sigma^\kappa = - x^\kappa \,\, ,
\end{align}
\begin{align}
&\bi{\sigma}{\alpha}{\kappa} = - \delta^\alpha_{\,\, \kappa}  + \frac{1}{3} {R_{\beta \kappa\gamma}}^\alpha x^\beta x^\gamma 
\nonumber\\
&\bi{\sigma}{\alpha}{\beta} = \delta^\alpha_{\,\, \beta}+ \frac{1}{3} {R_{\epsilon\beta\gamma}}^\alpha x^\epsilon x^\gamma 
\nonumber\\
&\bi{\sigma}{\kappa}{\alpha} = - \delta^\kappa_{\,\, \alpha} 
\nonumber\\
&\bi{\sigma}{\kappa}{\lambda} = \delta^\kappa_{\,\, \lambda} + \frac{1}{3} {R_{\alpha\lambda\beta}}^\kappa x^\alpha x^\beta \,\, ,
\end{align}
\begin{align}
&\bi{H}{\alpha}{\kappa} = \delta^\alpha_{\,\, \kappa} 
\nonumber\\
&\bi{K}{\alpha}{\kappa} = \delta^\alpha_{\,\, \kappa} + \frac{1}{3} {R_{\beta\kappa\gamma}}^\alpha x^\beta x^\gamma \, ,
\end{align}
showing that all these quantities only deviate from their flat spacetime counterparts, (\ref{flatbi1}-\ref{flatbi2}), at order $k(l/r)^2$. Other geometrical quantities in (\ref{UpPidef}) are the $n$-potential function $\langle n \rangle$, the volume element and the $w$-vector. The first two can be similarly shown to deviate from their flat spacetime counterparts by terms proportional to the curvature tensor. 

The $w$-vector, on the other hand, deserves a special consideration for it is not exclusively related to the curvature. For a straight line $\ell$ in the flat spacetime, it is clear that $w^a$ can be chosen as the covariant extension of $u^k = v^k$, that is, the vector field that coincides with $u^k$ on $\ell$ and satisfies $\nabla_a w^b = 0$ everywhere. If $\ell$ is taken as a general (non-straight) line, so that the orthogonal hypersurfaces $\Sigma(s)$ are not parallel to each other, then $w^a$ will not be a constant vector field. Let $z^\mu(s)$ be the trajectory of $\ell$ in the Riemann normal coordinate system centered at $z^\mu(0) = 0$, with temporal vector $(e_0)^\mu = v^\mu(0)$. Using the Taylor expansion in the proper length $s$, we can write
\begin{equation}
z^\mu(s) = (e_0)^\mu s + \frac{1}{2} a^\mu(0) s^2 \,\, ,
\end{equation}
where $a^\mu$ denotes the covariant acceleration $v^\nu \nabla_\nu v^\mu$. The hypersurface $\Sigma_{\delta s}$ at the infinitesimal proper time $\delta s$ is formed by all geodesics starting at $z^\mu(\delta s)$ orthogonal to $u^\mu(\delta s)$. Since $u^\mu$ only differs from $v^\mu$ by terms involving the curvature tensor, we may consider that $\Sigma_{\delta s}$ is actually formed by geodesics orthogonal to $v^\mu(\delta s)$ instead. Let $q^\mu(s)$ be a vector at $z^\mu(s)$ orthogonal to $v^\mu(s)$, so that the set of all $q^\mu(s)$ form a 3-dimensional vector subspace at $z^\mu(s)$. At $\delta s$, we have $0 = q^\mu(\delta s) v_\mu (\delta s) \approx \eta_{\mu\nu} q^\mu v_\nu = q^0 - \vec q \cdot \vec a \delta s$, where the curvature term in $g_{\mu\nu}(z(\delta s)) \approx \eta_{\mu\nu}$ was dropped. In this equation we are writing $q^\mu = (q^0, \vec q)$, in which $\vec q = (q^1, q^2, q^3)$, and $a^\mu = a^\mu(0) = (0, \vec a)$. The dot denotes the usual Euclidean scalar product, $\vec q \cdot \vec a = \sum_{i=1}^3 q^i a^i$. Thus, we have $q^\mu(\delta s) = ( \vec q \cdot \vec a \delta s, \vec q)$. A geodesic starting at $z^\mu(\delta s)$ and tangent to $q^\mu(\delta s)$, denoted by $x^\mu(\lambda)$, where $\lambda$ is any affine parameter, can be given approximately by
\begin{equation}
x^\mu(\lambda) = z^\mu(\delta s) + q^\mu(\delta s) \lambda - \frac{1}{3} \lambda^2 {R_{\nu\rho\sigma}}^\mu z^\nu(\delta s) q^\rho(\delta s) q^\sigma(\delta s) \, ,
\end{equation}
in which the last term, containing the curvature tensor, can be dropped. The $w$-vector can then be chosen at $\Sigma_0$ as the vector field satisfying
\begin{equation}
w^\alpha(0, \vec x(\lambda)) \delta s = (e_0)_\mu x^{\mu}(\lambda) (e_0)^\alpha
\end{equation}
for infinitesimal $\delta s$. The $x^\mu(\lambda)$ in the right-hand side of the equation has to be evaluated up to first order in $\delta s$, yielding $(e_0)_\mu x^{\mu}(\lambda) (e_0)^\alpha = (\delta s + \lambda  \vec q \cdot \vec a \delta s) (e_0)^\alpha$; while the argument of $w^\alpha$ in the left-hand side can be evaluated to the lowest order, $\vec x(\lambda) = \lambda \vec q$. It then follows that
\begin{equation}
w^\alpha (0, \vec x) = (1 + \vec x \cdot \vec a ) (e_0)^\alpha \,\, ,
\end{equation}
providing an expression for the $w$-factor for an arbitrarily specified curve $\ell$, up to curvature corrections. In our case, nonetheless, $\ell$ is the worldline of the center of mass and its 4-acceleration is given by equation (\ref{worldeq}), where it is clear that $\vec a$ necessarily contains a curvature factor (from the gravitational force and torque), that implies that its order of magnitude has a factor $\sim k/r^2$. Therefore the $w$-factor can be approximated as
\begin{equation}
w^\alpha \approx (e_0)^\alpha = \delta^\alpha_{\,\, 0} \,\, ,
\end{equation}
as if they were evaluated in a exactly flat spacetime (where $\ell$ would be exactly straight). 

Now that we have justified our claim that the multipole moments can be evaluated as if the spacetime were flat (under our approximations), let us make one assumption on the kind of bodies that shall be considered: from now on, we shall focus on {\sl articulated} bodies. More precisely, the assumption is that, at any instant, the {\sl internal configuration state} of the body can be described by a finite number of continuous parameters. In other words, there should exist an $n$-dimensional manifold $\ca I$, called the {\sl internal configuration space} of the body, such that there is a one-to-one correspondence between any point $q \in \ca I$ and a particular internal configuration state of the body (i.e., a specification of the instantaneous geometry of the body). Note that $\ca I$ is entirely external to the spacetime given that the instantaneous internal state of the body can be described in flat spacetime, which is a maximally symmetric space and contains no information about the structure of the real spacetime.

As an illustration, consider Wisdom's tripod in a 3-dimensional Euclidean space \cite{Wisdom}. It is defined by a ``pivot'' body, of proper mass $m_0$, connected to three other bodies, of equal proper masses $m_1$, by straight massless rods of length $l$. These three rods form equal angles to each other, defining a symmetrical tripod. Define the direction of the symmetry axis by a unit vector $\hat n_1$, specify the angle $\alpha$ that any of the rods makes with this axis, and let $\hat n_2$ (orthogonal to $\hat n_1$) indicate the orientation of the tripod around the symmetry axis. The parameters that can be controlled are $l$ and $\alpha$, while $\hat n_1$ and $\hat n_2$ are determined by the initial conditions and the dynamics of the tripod. Hence, if $l > 0$ and $0 < \alpha < \pi$, the internal space $\ca I$ of this body is diffeomorphic to $\mathbb R^+ \times (0,\pi)$. If we specify the routine of internal motions by giving the functions $l(s)$ and $\alpha(s)$, then the energy-momentum tensor of the body is determined on each hypersurface $\Sigma_s$, which can be used to evaluate the quadrupole and octupole moments at each ``time'' $s$. In Newtonian mechanics, the mass distribution will depend solely on the instantaneous configuration state $(l, \alpha)$, the momentum distribution will also depend on the ``velocities'' $(\dot l, \dot \alpha)$, where the dot denotes differentiation with respect to $s$, and the tensions may also depend on the ``accelerations'' $(\ddot l, \ddot \alpha)$; however, in relativity all these dynamical quantities will be mixed up in the momentum-energy tensor, so that we can only say that $T^{\mu\nu}$ will typically depend\footnote{If physics were still Newtonian in the local rest frame of any infinitesimal portion of the matter, this conclusion would follow naturally. Otherwise, this proposed functional dependence for the energy-momentum tensor may be seen as an additional assumption of the theory.} on $(l, \alpha, \dot l , \dot \alpha, \ddot l, \ddot \alpha)$.

Generally, in order to fully characterize the prescribed routine of shapes assumed by the body, we may specify a corresponding curve in the configuration space, $\gamma: \mathbb R \rightarrow \ca I$. In an arbitrary coordinate system $\{ \beta^\mu \}$ for $\ca I$, the curve can be described by the functions $\beta^\mu(s)$. The quadrupole and octupole moments are, supposedly, functions of the $\beta$'s, $\dot \beta$'s and $\ddot \beta$'s,
\begin{equation}
J = J ( \beta^1, \ldots \beta^n, \dot \beta^1,  \ldots \dot \beta^n, \ddot \beta^1,  \ldots \ddot \beta^n) \,\, ,
\end{equation}
where $J$ may represent either $J^{\mu\nu\rho\sigma}$ or $J^{\mu\nu\rho\sigma\lambda}$. For a given curve $\gamma$, the $J$'s become\footnote{Note that we are slightly abusing of the notation by using the same name for objects of different functional dependence, i.e., the $J$'s as functions of the point $z(s)$ (in the worldline $\ell$) or as functions of the configuration coordinates $\beta^i(s)$ and their derivatives. Since the arguments of the functions will be explicitly indicated, it should be clear which functional form is being used.} implicit functions of the parameter $s$,
\begin{equation}
J (z(s))= J ( \beta^1(s)  \ldots, \dot \beta^1(s) \ldots , \ddot \beta^1(s), \ldots) \, ,
\end{equation}
which can be used to solve for the worldline of the center of mass, $\ell$, according to the previous section. We may generalize this by considering that the $J$'s can also have an explicit parameter dependence
\begin{equation}
J (z(s)) = J ( \beta(s), \dot \beta(s), \ddot \beta(s), s) \,\, ,
\end{equation}
where $\beta$ denotes the full string of $\beta^i$, for $i = 1 \ldots n$, and analogously for $\dot \beta$ and $\ddot \beta$. The introduction of this explicit $s$ dependence can be useful in dealing with the case where the correspondence between points in $\ca I$ and the basic geometrical shape of the body change with time. For example, consider a body initially shaped like a three-dimensional ellipsoid, described by the lengths of its semi-axis $\beta^1$, $\beta^2$ and $\beta^3$, eventually morphing into a three-dimensional parallelepiped, with these same three parameters now representing the length of its edges (i.e., the parameter $s$ can be viewed as a {\sl homotopy} parameter of the continuous transformation that turns an ellipsoid into a rectangular parallelepiped).

Let us make a quick point on the parameterization of $\gamma$. In principle, we can take any physically meaningful parameter that can be diffeomorphically mapped to $s$, such as the time $\tau$ as measured by an observer moving instantaneously parallel to the dynamical velocity $u^k$. Note that the difference, in the case of $\tau$, does not need to be accounted since it affects the equations of motion only at a negligible order of magnitude, as can be seen by the chain rule
\begin{equation}
\frac{\delta}{ds} = \frac{d\tau}{ds} \frac{\delta}{d\tau} = u^k v_k \frac{\delta}{d\tau} \approx \left( 1 -  2 \epsilon^{kl} \epsilon_{rl} u_k u^r + \cdots \right) \frac{\delta}{d\tau} \,\, ,
\end{equation}
where $\epsilon^{kl}$ is the right-hand side of equation (\ref{uvmis}). Since it includes the curvature tensor, in the evaluation of the $J$'s one can simply approximate
\begin{equation}
\frac{\delta}{ds} \approx \frac{\delta}{d\tau}  \,\, ,
\end{equation}
showing that it is meaningless to make a distinction between derivatives in $s$ or in $\tau$. Also, if the total time for which one is interested in evolving the dynamical quantities is not too large, these parameterizations are really indistinguishable,
\begin{align}
\tau(s) &= \int_0^s u^k(\bar s) v_k(\bar s) d\bar s = s -2 \int_0^s \epsilon^{kl} \epsilon_{rl} u_k u^r d\bar s + \cdots 
\nonumber\\
&\approx s - 2s \ca O (\epsilon^2) \, ,
\label{stau}
\end{align}
that is, if $s \ca O (\epsilon^2)$ is sufficiently small, then $\tau(s) \approx s$. Thus, for simplicity, we will use $s$ as the standard parameter of $\gamma$.

Finally, let us consider an important detail that has not been emphasized in the discussion above. By using this local mapping to a flat spacetime in a neighborhood of each point $z(s)$, we have seen that it was possible to evaluate the energy-momentum tensor on $\Sigma_s$ which, consequently, determined the quadrupole and octupole moments at $z(s)$. But how to connect this mapping for different points $z(s)$? That is, how the components $J^{\mu\nu\rho\sigma}$ obtained in the Minkowski spacetime translate into a tensor $J^{klqr}$ in the true spacetime? Let us discuss a very natural proposal for establishing this connection. Take the initial position of the center of mass to be $z(s_0)$ and the initial dynamical velocity to be $u^k(s_0)$. Let $\{(e_\mu)^k\}$ be any tetrad at $z(s_0)$ with $(e_0)^k = u^k(s_0)$. At this point, $z(s_0)$, consider the correspondence between curved and flat spacetimes induced by the inverse exponential mapping from some (open) neighborhood $\widetilde \Sigma_{s_0}$ of $\Sigma_{s_0} \cap W$ to $T_{z(s_0)}\ca M$, and set global inertial (Cartesian) coordinates in $T_{z(s_0)}\ca M$ such that each base vector $\partial/\partial x^\mu$ is mapped to the corresponding $(e_\mu)^k$. It is then natural to assume that, at this point, one should have
\begin{equation}\label{barquad}
J^{klqr}(z(s_0)) = \bar J^{\mu\nu\rho\sigma}(z(s_0)) (e_\mu)^k (e_\nu)^l (e_\rho)^q (e_\sigma)^r \, ,
\end{equation}
and analogously for $J^{klqrm}(z(s_0))$. The bar is only to make clear that the $\bar J$'s are evaluated in the flat spacetime. As the time passes, we can solve the dynamics of the body in the flat spacetime so as to get the value of its energy-momentum tensor at any future time $t > s_0$, which allows one to evaluate the multipole moments, $\bar J^{\mu\nu\rho\sigma}(\beta(t), \dot \beta(t), \ddot \beta(t), t)$ and $\bar J^{\mu\nu\rho\sigma\lambda}(\beta(t), \dot \beta(t), \ddot \beta(t), t)$, in the inertial coordinate frame by using the information of the energy-momentum on the section $(x^0 = t, x^1, x^2, x^3)$. 
The assumption here is that, as long as $t$ is not too large (see formula (\ref{stau})), then expression (\ref{barquad}) can be extended without modifications to an arbitrary point $z(s)$ by evaluating the $\bar J$'s at the time $t = \tau(s) \approx s$ (where $\tau(s_0) = s_0$),
\begin{align}
&J^{klqr}(z(s)) = 
\nonumber\\
&\bar J^{\mu\nu\rho\sigma}(\beta(\tau(s)), \dot \beta(\tau(s)), \ddot \beta(\tau(s)), \tau(s)) (\bar e_\mu)^k (\bar e_\nu)^l (\bar e_\rho)^q (\bar e_\sigma)^r
\label{barquad2}
\end{align}
and analogously for $J^{klqrm}(z(s))$, where the tetrad $\{(\bar e_\mu)^k\}$ at $z(s)$ is obtained by any reasonable transportation of $\{(e_\mu)^k\}$ from $z(s_0)$ along $\ell$. For instance, we can consider the M-transport 
\begin{equation}
\frac{\delta (\bar e_\mu)^k}{ds} = (\dot u^k u_l - u^k \dot u_l)(\bar e_\mu)^l \,\, ,
\end{equation}
or the Fermi transport or the parallel transport, and they should all agree within our approximations. This completes the discussion on the basic idea behind the prescription of the quadrupole and octupole moments, which is going to be central to the study of the swimming phenomenon.

\section{Swimming deviation formula}
\label{sec:Swimming}

In the this section we will consider the theory of swimming in spacetime, investigating the possibility that a body could move through the space in a way resembling the motion of swimmers in non-turbulent viscous fluids --- or, more precisely, at low Reynolds number\footnote{The Reynolds number is defined as $\mathscr{R} = \rho u l/\mu$, where $\rho$ is the mass density of the fluid, $u$ is the velocity of the fluid with respect to the body, $l$ is the characteristic length scale of the body and $\mu$
is the (dynamical) viscosity of the fluid.}. 
We are thus interested in finding a regime where the effective displacement of the center of mass does not depend on how fast the internal motions happen, but only on the geometric phase associated with the sequence of shapes assumed by the body. As we are going to see, this is not very typical of the motion of free bodies in general relativity, but can take place in particular scenarios and settings. Our formulas, nevertheless, are not limited to this special regime and can describe the dynamics of general small articulated bodies, free falling in curved spacetimes, performing internal motions, as long as the observation time is not too long.

Let the internal configuration space $\ca I$ be $n$-dimensional, covered by coordinates $(\beta^1, \beta^2, \ldots \beta^n)$. Let there be $d$ dynamical variables $(q^1, q^2, \ldots q^d)$ describing the global configuration of the body -- they may correspond, for instance, to the three spatial coordinates $(x^1, x^2, x^3)$ of the center of mass, the three spatial components of the dynamical velocity $(u^1, u^2, u^3)$, the spatial components of the spin $(S^1, S^2, S^3)$, the proper mass $M$, or any other dynamical attribute associated with the body (these components being associated with a suitable coordinate system, such as a Riemann normal coordinate system around the initial position, aligned with the initial velocity of the center of mass). Let $q_0(t)$ be the solution for the dynamical quantities when the body is kept ``rigid'', i.e., with fixed configuration coordinates $\beta_0$. Define the {\sl deviation} of the dynamical quantity $q^i$, denoted by $\zeta^i$, as the difference between the general solution and the rigid body solution,
\begin{equation}
\zeta^i(t) = q^i(t) - q_0^i(t) \,\, ,
\end{equation}
which is supposed to measure how the internal motions affect the motion of the center of mass. 

It is clear that when the body performs a certain sequence of internal motions, represented by a curve $\gamma : [0, t] \rightarrow \ca I$ in the internal configuration space, the dynamical equations can be solved to determine $\zeta(t)$. In other words, keeping all other parameters fixed (spacetime curvature, initial conditions, etc), the variations $\zeta^i$ are functions of the curve $\gamma$, $\zeta^i = f^i(\gamma)$. The swimming at ``low Reynolds number'' happens when the deviations $\zeta^i$ display a geometric phase behavior, depending only on the {\sl sequence} of internal motions. That is, the deviations $\zeta^i$ become functions of the {\sl image} of the curve only, $\zeta^i = f^i(\mathrm{Im}(\gamma))$.

We shall search for a regime in which the deviation variables satisfy a set of differential equations of the form
\begin{equation}\label{geophase}
\dot\zeta^i = \lambda^i_{\,\, a'} (\beta, \zeta) \dot\beta^{a'} \,\, ,
\end{equation}
where the dot continues to denote derivatives in the time $t$, there is an implicit summation over $a' = 1 \ldots n$, and $\beta$ and $\zeta$ are short notations for the strings of variables $(\beta^1, \beta^2, \ldots \beta^n)$ and $(\zeta^1, \zeta^2, \ldots \zeta^d)$, respectively. In this section, we shall use a different convention for the indices, where Latin letters will correspond to components of the dynamical ($q^i$) or deviation ($\zeta^i$) variables (assuming values from $1$ to $d$), and primed Latin letters will correspond to configuration coordinates $\beta^{a'}$ (assuming values from $1$ to $n$); the Einstein summation convention for repeated indices will be preserved. 

Let us show that, in such regime, the so described geometric-phase behavior is observed.
The space of deviations $\zeta^i$, denoted by $\ca Z$, is assumed to be a manifold diffeomorphic to $\mathbb R^d$. It is then convenient to define the {\sl state space} $\ca P$ of the problem as the Cartesian product $\ca P = \ca I \times \ca Z$, so that each point of it corresponds to a certain internal configuration and dynamical state of the body. Since $\ca Z$ has a vector space structure, $\ca P$ can be seen as the vector bundle $(\ca P, \ca I, \rho_1)$, where $\ca I$ is the base space and the projection of the bundle is the projection on the first component of the product $\ca I \times \ca Z$; this means that for each internal configuration state in $\ca I$ there is a fiber attached to it representing the dynamical space of the body. We shall use the natural isomorphism $T_p\ca P \sim T_\beta \ca I \oplus T_\zeta \ca Z$, where $p = (\beta, \zeta)$ is a point in $\ca P$, to represent vectors in the state space; in the component form it will generally read $V = (v^1, \ldots v^n; \eta^1, \ldots \eta^d)$, succinctly written as $V = (v^{a'}; \eta^i)$. For an arbitrary vector $v$ at $T_\beta \ca I$, define the lifted vector $V$ in $T_{(\beta, \zeta)} \ca P$ as
\begin{equation}
V = (v^{a'}; \lambda^i_{\,\, {a'}} (\beta, \zeta) v^{a'}) \,\, ,
\end{equation}
which can be alternatively expressed as a linear map $\Lambda : T_\beta \ca I \rightarrow T_{(\beta, \zeta)} \ca P$ 
\begin{equation}
V = \Lambda v \,\, .
\end{equation}
Note that this map $\Lambda$ can be used to lift a vector field $v$ in $\ca I$ to a vector field $V$ in $\ca P$. Also, any curve $\gamma : \mathbb R \rightarrow \ca I$ can be lifted to a curve $\Gamma : \mathbb R \rightarrow \ca P$ such that, given a point $p \in \ca P$ in the fiber at $\gamma(0)$, then $\Gamma(t)$ is the unique curve in $\ca P$ satisfying $\Gamma(0) = p$ and
\begin{equation}
\Gamma'(t) = \Lambda \gamma'(t) \,\, .
\end{equation}
Observe that if we write $\Gamma(t)$ in the coordinates $\{ \beta, \zeta \}$ as $\Gamma(t) = (\beta^{a'}(t), \zeta^i(t))$, then the equation above translates to the system of differential equations
\begin{align}
\dot\beta^{a'}(t) &= \dot\gamma^{a'}(t) \\
\dot\zeta^i(t) &= \lambda^i_{\,\, {a'}} (\beta(t), \zeta(t)) \dot\gamma^{a'}(t) 
\end{align}
that can be immediately identified as the differential equation (\ref{geophase}). In other words, the curve $\Gamma(t)$ represents the precise solution of equation (\ref{geophase}) when the curve $\gamma(t)$ is traversed in the configuration space. Therefore, the projection of $\Gamma(t)$ by  $\rho_2$ (i.e., the projection on $\ca Z$) gives the exact values of the dynamical deviations $\zeta^i(t)$, with initial conditions $\zeta(0) = \rho_2(\Gamma(0)) = \rho_2(p)$, when the internal motions run along $\gamma(t)$.

\begin{figure}[!h]
\centering
\includegraphics[scale = 0.5]{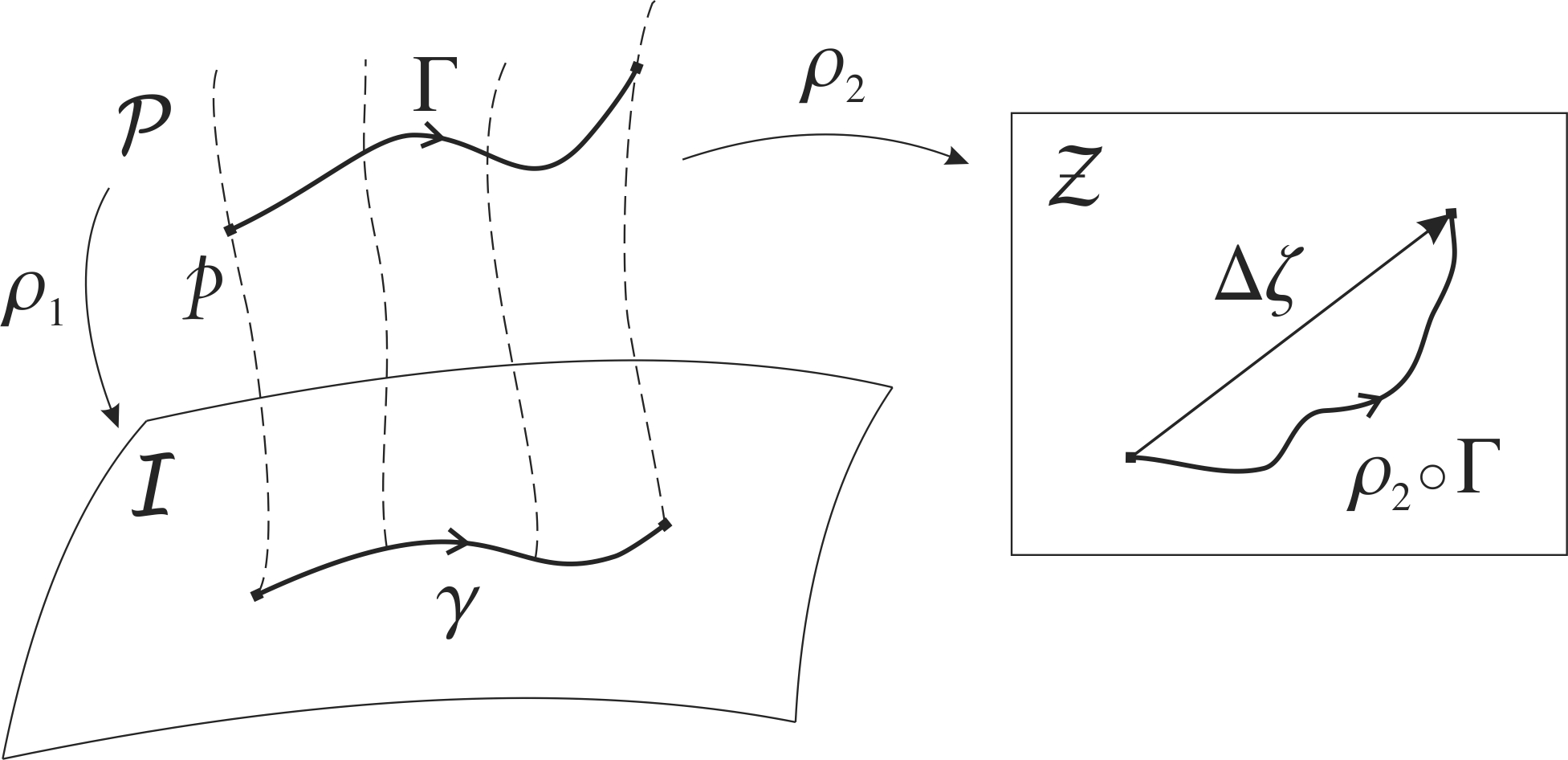}
\caption{The state space $\ca P$, the configuration space $\ca I$ and the dynamical space $\ca Z$ are shown (the dotted lines represent fibers of $\ca P$). A curve $\gamma$ in $\ca I$ is lifted to a curve $\Gamma$ in $\ca P$ starting at the point $p \in \ca P$. The projection $\rho_2$ maps $\Gamma$ to $\ca Z$, where the total deviation of the dynamical quantities $\Delta \zeta$ is the vector connecting the initial and final points of the projected curve.}
\label{FIGswimphase}
\end{figure}

Note that the total deviation of the dynamical quantities is given by the difference of $\rho_2 \circ \Gamma$ at the final point and the initial point $\Gamma(0) = p$ [see figure (\ref{FIGswimphase})]. Therefore, it is a function of the image of $\Gamma$ between the initial and final points (thence independent of its parametrization). In addition, the image of $\Gamma$ is a function of the image of $\gamma$, which can be seen as follows.
Consider a general reparametrization of $\gamma$ given by $\bar \gamma = \gamma \circ \chi$, where $\chi : \mathbb R \rightarrow \mathbb R$ is a smooth function. The chain rule implies that
\begin{equation}
\bar\gamma' = \gamma' \circ \chi' = \frac{d\chi}{dt} \gamma' \,\, ,
\end{equation}
Since $\Lambda$ is a linear map, we have
\begin{equation}
\bar\Gamma'(t) = \Lambda \bar\gamma'(t) = \frac{d\chi}{dt} \Lambda \gamma'(t) = \frac{d\chi}{dt} \Gamma'(t) \,\, ,
\end{equation} 
showing that the curve $\bar\Gamma$ lifted from $\bar\gamma$ is just a reparametrization of the curve $\Gamma$ lifted from $\gamma$, $\bar\Gamma =  \Gamma \circ \chi$.
Thus, as anticipated, the net deviation of the dynamical quantities is a function of the image of $\gamma$ and conseqiuently the dynamics of the problem is indeed of the geometrical phase kind.

Let us now derive a formula for the net deviation of the dynamical quantities when $\gamma$ is a small closed curve. More precisely, we shall limit ourselves to the case of a small rectangular cycle defined by two transversal smooth vector fields $u$ and $v$ in the configuration space $\ca I$. Let the parameters of the integral curves of the vector fields $u$ and $v$ be $x$ and $y$, respectively. Assume that $\gamma$ starts at the point $\beta_0$, then goes along the integral curve of $u$ by an small variation in the parameter $\Delta x$, then moves along the direction of $v$ by a parameter change $\Delta y$, then goes back in the direction of $u$ by a parameter $-\Delta x$, and finally moves along $v$ by an parameter $-\Delta y$. After going though the four sides of this ``parallelogram'', the curve will have returned to the initial point $\beta_0$ if and only if these two vector fields commute
\begin{equation}
[u, v] = 0 \,\, ,
\end{equation}
and this will be assumed to hold. The lifted curve, $\Gamma$, will be precisely the curve obtained by going through the corresponding parallelogram defined by the lifted vector fields $U = \Lambda u$ and $V = \Lambda v$. Since the lifted fields may not commute, $\Gamma$ may not be closed and there may be some non-vanishing net deviation of the dynamical quantities after the cycle is traversed. For the quantity $\zeta^i$, we have that, at lowest approximation, its variation is given by
\begin{equation}
\Delta \zeta^i = \Big. [U, V] \zeta^i \Big|_{(\beta_0, \zeta_0)} \Delta x \Delta y \,\, ,
\end{equation}
where $\Delta \zeta^i = \zeta^i (\Gamma(\Delta t)) - \zeta_0^i$. For simplicity, let us redefine the vector fields so as to include $\Delta x$ and $\Delta y$ implicitly in the respective vector
\begin{align}
u \Delta x &\mapsto u \\
U \Delta x &\mapsto U \\
v \Delta y &\mapsto v \\
V \Delta y &\mapsto V \,\, ,
\end{align}
so that the new fields can be seen as fields of infinitesimal vectors, and the formula above can be put in a slightly more appealing form
\begin{equation}
\Delta \zeta^i = [U, V] \zeta^i \,\, .
\end{equation}
If $\partial$ is the derivative operator associated with the coordinates $(\beta, \zeta)$, we have
\begin{equation}
\Delta \zeta^i = [U,V]^{a'} \partial_{a'} \zeta^i + [U,V]^j \partial_j \zeta^i = [U,V]^i \,\, ,
\end{equation}
where it was used that $\partial\zeta^i/\partial\beta^{a'} = 0$ and $\partial\zeta^i/\partial\zeta^j = \delta^i_{\,\, j}$. After some simple manipulations we get
\begin{equation}
\Delta \zeta^i = \lambda^i_{\,\, {a'}} [u,v]^{a'} + 2 \left( \partial_{a'} \lambda^i_{\,\, {b'}} + \lambda^j_{\,\, {a'}} \partial_j \lambda^i_{\,\, b'} \right) u^{[{a'}} v^{b']} \,\, .
\end{equation}
Thus, using $[u, v]^{a'} = 0$, we obtain
\begin{equation}
\Delta \zeta^i = 2 \left( \partial_{a'} \lambda^\mu_{\,\, b'} + \lambda^j_{\,\, {a'}} \partial_j \lambda^i_{\,\, b'} \right) u^{[{a'}} v^{b']} \,\, ,
\end{equation}
where the right-hand side is evaluated at $(\beta_0, \zeta_0)$. 

It is interesting to point out that if $\lambda$ in (\ref{geophase}) depended explicitly on the time $t$, then the motion would not display this geometric phase character, so that the net variation of the dynamical quantities would depend on $\gamma(t)$ and not only on $\mathrm{Im}(\gamma)$. Notwithstanding, the geometric phase character of the dynamics is restored in the limit of small evolution times, for we can approximate
\begin{equation}
\dot\zeta^i = \lambda^i_{\,\, {a'}} (\beta, \zeta, t) \dot\beta^{a'}  \approx \lambda^i_{\,\, {a'}} (\beta, \zeta, t_0) \dot\beta^{a'} + \left. \frac{\partial\lambda^i_{\,\, {a'}}}{\partial t} \right|_{ (\beta, \zeta, t_0)} \Delta t \dot\beta^{a'} 
\end{equation}
where $\Delta t = t - t_0$ is small during the whole integration (for instance, during the small time lapse of a fast cycle). Thus, one can obtain a fairly good approximation for the solution by considering the approximate differential equation
\begin{equation}
\dot\zeta^i \approx \lambda^i_{\,\, {a'}} (\beta, \zeta, t_0) \dot\beta^{a'}  \,\, ,
\end{equation}
which displays the desired geometrical-phase kind of dynamics we just studied. It is in this sense that we refer to ``small time'' in this problem.

Note that if $\dot\zeta$ had a different functional dependence on $\dot\beta$ or there were terms containing higher derivatives such as $\ddot\beta$, then the motion would (most likely) not be of the geometrical-phase kind, even in the limit of fast cycles. Indeed, for very fast cycles, $\dot\beta$ and $\ddot\beta$ would increase accordingly so as to contribute sensibly to the dynamics of the problem. Thus, we shall look for a particular scenario in which small bodies in general relativity could be well described by equations of the form (\ref{geophase}). 

Note that if the functions $\lambda$ were ``small'', then
\begin{equation}\label{SwimmingFormula}
\Delta \zeta^i \approx 2  \partial_{a'} \lambda^i_{\,\, b'} u^{[{a'}} v^{b']} \,\, ,
\end{equation}
where the term involving $\lambda^2$ were dropped. This implies that it is only the $\beta$-dependence of $\lambda$ that effectively matters in the final formula, so that $\lambda(\beta, \zeta)$ can be evaluated at the initial values of the dynamical quantities
\begin{equation}
\lambda^i_{\,\, {a'}}(\beta, \zeta) \mapsto  \lambda^i_{\,\, {a'}}(\beta, \zeta_0)  \,\, .
\end{equation}
This will turn out to be the case of our interest, since the functions $\lambda$ will contain factors of curvature and thence contribute with factors of order of magnitude $k (l/r)^2$ or higher, so that $\lambda^2$ would be of order $k^2 (l/r)^4$ or higher, negligible in our approximations.

The equation for the trajectory of the center of mass, given in (\ref{worldeq}), is of second order in $z^\mu(s)$. Since we will be mostly interested in fast cycles of internal motions, we will only have to solve these equations for a small time interval $s$ (assuming that the initial conditions are given at $s=0$). It is then convenient to take the local normal coordinate system centered at $z(0)$ (so that $z^\mu(0) = 0$) and oriented along the initial dynamical velocity $u^k(0)$ (so that $u^\mu(0) = (1,0,0,0)$). If the trajectory of the center of mass is described by $z^\mu(s)$, then $v^l \nabla_l v^k$ decomposes as
\begin{equation} 
v^\nu\nabla_\nu v^\mu = \frac{d^2 z^\mu}{ds^2} + \Gamma^\mu_{\,\, \nu\sigma}(z) \frac{dz^\nu}{ds} \frac{dz^\sigma}{ds}  \,\, ,
\end{equation}
where $\Gamma^\mu_{\,\, \nu\sigma}(z)$ can be approximated as
\begin{equation} 
\Gamma^\mu_{\,\, \nu\sigma}(z) = \frac{2}{3}  {R_{\rho(\nu\sigma)}}^\mu(0) z^\rho + \frac{1}{2}\partial_\rho\partial_\kappa \Gamma^\mu_{\,\, \nu\sigma}(0) z^\rho z^\kappa \,\, ,
\end{equation}
in which $\partial_\rho\partial_\kappa \Gamma^\mu_{\,\, \nu\sigma}(0)$ can be expressed in terms of the first derivative of the curvature at the origin, $\nabla_\kappa {R_{\rho\sigma\nu}}^\mu(0)$. Thus, equation (\ref{worldeq}) becomes
\begin{align}
\ddot z^\mu &= - \Gamma^\mu_{\,\, \nu\sigma}(z) \dot z^\nu \dot z^\sigma + \frac{1}{M} \left( F^\mu - F^\nu u_\nu u^\mu \right) 
\nonumber\\
&+ \frac{1}{M^2} S^{\mu\nu} \dot F_\nu + \frac{1}{M} \dot Q^{\mu\nu} u_\nu \, ,
\label{ddotzgen}
\end{align}
where the Christoffel symbols associated with the derivatives of the force and torque could be dropped due to our approximations. For the sake of completeness, consider the differential equations corresponding to the time evolution of other relevant dynamical quantities
\begin{align}
&v^l \nabla_l u^k = \frac{1}{M} \left( F^k - F^l u_l u^k \right)
\nonumber\\
&v^l \nabla_l M =  F^l u_l 
\nonumber\\
&v^r \nabla_r S^{kl} = 2 M u^{[k} v^{l]} + Q^{kl}
\nonumber\\
&v^l \nabla_l (\bar e_\mu)^k = \frac{2}{M}  F^{[k} u^{l]} (\bar e_\mu)_l \,\, .
\end{align}
Note that, integrating these equations to the lowest order in time we obtain something of the form
\begin{equation}
\zeta(s) \approx \zeta(0) + \alpha s \,\, ,
\end{equation}
where $\alpha$ is a quantity that necessarily contains a factor of the magnitude $k/r^2$. Hence, considering our approximations, all the dynamical quantities in the right-hand side of equation (\ref{ddotzgen}) can be approximated by their initial values
\begin{align}
&M \mapsto M_0 
\nonumber\\
&u^\mu \mapsto \delta^\mu_{\,\, 0} 
\nonumber\\
&S^{\mu\nu} \mapsto S_0^{\mu\nu} 
\nonumber\\
&(\bar e_\nu)^\mu \mapsto \delta^\mu_{\,\,\nu} \, ,
\end{align}
as long as $s$ is not too big. Also, since the kinematical velocity $\dot z^\mu$ only differs from the dynamical velocity $u^\mu$ by a quantity containing the curvature, we may also take
\begin{equation}
\dot z^\mu \mapsto \delta^\mu_{\,\, 0}
\end{equation}
in the right-hand side. Thence, equation (\ref{ddotzgen}) can be simplified to
\begin{equation}\label{ddotzapp}
\ddot z^\mu = - \Gamma^\mu_{\,\, 00}(z) + \frac{1}{M_0} \left( F^\mu - F^0 \delta^\mu_{\,\, 0} \right) + \frac{1}{M_0^2} S_0^{\mu\nu} \dot F_\nu + \frac{1}{M_0} \dot Q^{\mu0} \, ,
\end{equation}
and, in addition, the force and torque can be given by
\begin{align}
F^\mu &= \frac{1}{2} S_0^{\nu\rho} {R^\mu}_{0\nu\rho} + \frac{1}{6} \bar J^{\nu\rho\sigma\lambda}\nabla^\mu R_{\nu\rho\sigma\lambda}
\nonumber\\
Q^{\mu0} &= \frac{4}{3} \bar J^{\nu\rho\sigma[\mu} {R_{\nu\rho\sigma}}^{0]} \, +
\nonumber\\
& + \frac{1}{3} \bar J^{\nu\rho\sigma\lambda[\mu} \left( 3 \nabla_\nu {R_{\rho\sigma\lambda}}^{0]} + 2 \nabla_\lambda {R_{\nu\rho\sigma}}^{0]} \right) \, ,
\end{align}
where it was used that $J^{\mu\nu\rho\sigma} = \bar J^{\mu\nu\rho\sigma}$ and $J^{\mu\nu\rho\sigma\lambda} = \bar J^{\mu\nu\rho\sigma\lambda}$ since the non-spinning basis $\bar e_\mu$ coincides with the coordinate basis $\partial_\mu$. Note also that the octupole contribution to the force was omitted, which is due to the fact that it depends on the second derivative of the curvature tensor, assumed to be of order $k/r^4$.

The $z^\mu$ appearing in the right-hand side of equation (\ref{ddotzapp}) can be approximated using a similar argument as the one employed for the dynamical quantities. By equation (\ref{vapprox2}), we conclude that
\begin{equation}
z^\mu(s)  \approx s \delta^\mu_{\,\, 0} + \alpha^\mu s \,\, ,
\end{equation}
where $\alpha^\mu$ are quantities involving factors of curvature $\sim k/r^2$. Hence, the $z^\mu$ that appears in the arguments of $\Gamma^\mu_{\,\, 00}(z)$ and the curvature (and its derivatives) can be approximated by $z^\mu(s) = (s, 0, 0, 0)$.

The equation for the spatial position of the center of mass, $z^i(s)$, is then given by
\begin{equation}\label{ddotzapp2}
\ddot z^i = - \Gamma^i_{\,\, 00}(z) + \frac{1}{M_0} F^i + \frac{1}{M_0^2} S_0^{ij} \dot F_j + \frac{1}{M_0} \dot Q^{i0}
\end{equation}
or, alternatively,
\begin{equation}\label{dotzapp2A}
\frac{d}{ds} \left[ \dot z^i - \frac{1}{M_0^2} S_0^{ij}  F_j - \frac{1}{M_0}  Q^{i0} \right] =  - \Gamma^i_{\,\, 00}(z) + \frac{1}{M_0} F^i \,\, .
\end{equation}
But this equation still does not have the desired form (\ref{geophase}). 

At this point it is convenient to assume that $\bar J$ is linear in $\ddot \beta$, that is,
\begin{equation}
\bar J(\beta, \dot\beta, \ddot\beta, s) = \bar J(\beta, \dot\beta, 0, s) + \left. \frac{\partial \bar J}{\partial \ddot\beta} \right|_{(\beta, \dot\beta, 0, s)} \ddot\beta \, .
\end{equation}
This is motivated by the fact that, in Newtonian mechanics, the $\ddot \beta$'s in the $\bar J$'s typically come from the tension contributions in the energy-momentum tensor, which usually appear linearly in the stress tensor. Since a boost by $\dot \beta$ cannot introduce extra $\ddot \beta$ in the stress-energy tensor, this should still be true in special relativity, as long as the local physics in the instantaneous rest frame of each infinitesimal piece of matter is essentially Newtonian -- if this is not true, then we shall take this special form of $\bar J$ as an additional assumption. Accordingly, the following quantities can be written as
\begin{align}
&\frac{1}{M_0} F^i = \phi^i(\beta, \dot\beta, s) + \widetilde\phi^i_{\,\, a'} (\beta, \dot\beta, s) \ddot\beta^{a'}
\nonumber\\
&\frac{1}{M_0^2} S_0^{ij}  F_j + \frac{1}{M_0}  Q^{i0} = \psi^i(\beta, \dot\beta, s) + \widetilde\psi^i_{\,\, a'} (\beta, \dot\beta, s) \ddot\beta^{a'} \,\, ,
\end{align}
where the functions $\phi$, $\widetilde\phi$, $\psi$ and $\widetilde\psi$ depend implicitly on the initial conditions. The components of the spatial acceleration $\ddot z^i$ can be generally cast in the form
\begin{align}\label{ddotzapp3}
&\frac{d}{ds} \left[ \dot z^i  - \widetilde\phi^i_{\,\, a'} (\beta, \dot\beta, s) \dot\beta^{a'} - \psi^i(\beta, \dot\beta, s) - \widetilde\psi^i_{\,\, a'} (\beta, \dot\beta, s) \ddot\beta^{a'} \right] 
\nonumber\\
& = \epsilon^i(s) + \phi^i(\beta, \dot\beta, s) - \frac{d\widetilde\phi^i_{\,\, a'} }{ds} \dot\beta^{a'} \,\, ,
\end{align}
where $\epsilon^i(s) = - \Gamma^i_{\,\, 00}(z)$. Note that the quantity inside the square brackets in the left-hand side corresponds to a first order differential equation in $z^i$ that differs from equation (\ref{geophase}) only by the term with $\ddot\beta$ and the possible non-linearity in $\dot\beta$. If we integrate equation (\ref{ddotzapp3}) in time, it is conceivable that the right-hand side could be neglected for very short times of evolution,
\begin{align}
&\dot z^i  - \widetilde\phi^i_{\,\, a'} (\beta, \dot\beta, s) \dot\beta^{a'} - \psi^i(\beta, \dot\beta, s) - \widetilde\psi^i_{\,\, a'} (\beta, \dot\beta, s) \ddot\beta^{a'}   =
\nonumber\\
& \qquad = \int_0^s (\cdots) ds' \approx 0 \, ,
\end{align}
which could be true if the right-hand side of (\ref{ddotzapp3}) depended only on $\beta$, $\dot\beta$ and $s$, and not on superior derivatives of $\beta$. However, the term involving the time derivative of $\widetilde\phi$ can be expressed as
\begin{equation}
\frac{d\widetilde\phi^i_{\,\, a'} }{ds} \dot\beta^{a'} = \left( \frac{\partial \widetilde\phi^i_{\,\, a'}}{\partial \beta^{b'}} \dot\beta^{b'} + \frac{\partial \widetilde\phi^i_{\,\, a'}}{\partial \dot\beta^{b'}} \ddot\beta^{b'} + \frac{\partial \widetilde\phi^i_{\,\, a'}}{\partial s} \right) \dot\beta^{a'} \,\, ,
\end{equation}
and we can clearly see that it may depend\footnote{Note that one could have tried to integrate $\widetilde\phi^i_{\,\, a'} \ddot\beta^{a'}$ into the right-hand side of equation (\ref{ddotzapp3}) by finding a function $\widehat\phi^i(\beta, \dot\beta, s)$ such that $\partial \widehat\phi^i/\dot\beta^{a'} = \widetilde\phi^i_{\,\, a'}$. However, that would only be possible if $\partial \widetilde\phi^i_{\,\, a'} / \partial \dot\beta^{b'} - \partial \widetilde\phi^i_{\,\, b'} / \partial \dot\beta^{a'} = 0$, and there is no physical reason for that being true.} on $\ddot\beta$. Nonetheless, this $\ddot\beta$-dependent term is absent if we ask for the condition
\begin{equation}\label{HnewbarJ}
\frac{\partial}{\partial\dot\beta^{a'}} \left( \frac{\partial \bar J}{\partial\ddot\beta^{b'} } \right) = 0 \, ,
\end{equation}
which means that the terms in the energy-momentum tensor $T^{\mu\nu}$ involving the ``accelerations'' $\ddot\beta$ would not depend on the ``velocities'' $\dot\beta$. This is fairly common in Newtonian problems, where the tensions tend to assume the form
\begin{equation}\label{HnewbarJ2}
\tau \sim \kappa(\beta) + \theta(\beta) \dot\beta \dot\beta + \omega (\beta) \ddot\beta
\end{equation}
and it can be easily verified that this form is invariant by a (velocity-independent) change of coordinates $\beta \rightarrow \beta'(\beta)$. Under this assumption, we can write equation (\ref{ddotzapp3}) as
\begin{align}
&\frac{d}{ds} \left[ \Delta \dot z^i  - \widetilde\phi^i_{\,\, a'}(\beta, s)\dot\beta^{a'} - \psi^i(\beta, \dot\beta, s) - \widetilde\psi^i_{\,\, a'}(\beta, s)\ddot\beta^{a'} \right] =
\nonumber\\
&=  \phi^i(\beta, \dot\beta, s) - \left( \frac{\partial \widetilde\phi^i_{\,\, a'}}{\partial \beta^{b'}} \dot\beta^{b'} + \frac{\partial \widetilde\phi^i_{\,\, a'}}{\partial s} \right) \dot\beta^{a'}  \,\, ,
\end{align}
where the arguments of $\widetilde\phi$ and $\widetilde\psi$ were accordingly reduced to $(\beta, s)$.

It seems difficult to proceed much further from this point if we do not make any assumptions on the $\dot\beta$'s. Clearly, since they are mainly related with the velocities, it seems reasonable to assume that they are bounded. For example, if $(l, \alpha)$ are the radial coordinates of point moving on a 2-dimensional plane, then its velocity would be given by $\vec v = (\dot l , l \dot\alpha )$, which puts a restriction of the kind $\dot l \lesssim 1$ and $\dot\alpha \lesssim l^{-1}$. We shall, nonetheless, go beyond and assume that the $\dot\beta$'s are ``small'' quantities. This is supposed to ensure that the internal motions are slow enough so as to be essentially Newtonian, such that relativistic-related non-linear terms in the velocities could be neglected in the equations. Note that this is essential to the tenability of hypothesis (\ref{HnewbarJ}), whose underlying argument is based in (\ref{HnewbarJ2}), that is only reasonable in the Newtonian regime (in fact, a boost would necessarily introduce non-linear velocity factors in the tensions). Thus, we shall assume that the velocities are small and that the tensions are much smaller that the mass density scale of the body, so that the dynamics of the internal motions can be treated in a purely Newtonian way.

Let $z_0(s)$ be the trajectory of the rigid body (i.e., when $\beta(s) = \beta_0$), but starting with the same initial conditions $z_0^\mu(0) = 0$ and $\dot z_0^\mu(0) =\delta^\mu_{\,\, 0}$. Its second derivative satisfies the equation
\begin{equation}
\frac{d}{ds} \left[ \Delta \dot z_0^i  - \psi^i(\beta_0, 0, s) \right] =  \epsilon^i(s) + \phi^i(\beta_0, 0, s) \,\, ,
\end{equation}
so that the deviation $\Delta z^i(s) = z^i(s) - z_0^i(s)$ satisfies
\begin{align}
&\frac{d}{ds} \left[ \Delta \dot z^i  - \widetilde\phi^i_{\,\, a'}(\beta,  s)\dot\beta^{a'} - \psi^i(\beta, \dot\beta, s) \right. \nonumber\\
&\qquad\qquad \left. +\, \psi^i(\beta_0, 0, s) - \widetilde\psi^i_{\,\, a'}(\beta, \dot\beta, s)\ddot\beta^{a'} \right] =
\nonumber\\
&=  \phi^i(\beta, \dot\beta, s) - \phi^i(\beta_0, 0, s) - \left( \frac{\partial \widetilde\phi^i_{\,\, a'}}{\partial \beta^{b'}} \dot\beta^{b'} + \frac{\partial \widetilde\phi^i_{\,\, a'}}{\partial s} \right) \dot\beta^{a'} \,\, .
\end{align}
Expanding $\phi$ and $\psi$ in Taylor and keeping first order terms in $\Delta\beta := \beta - \beta_0$ but second order terms in $\dot\beta$ we get
\begin{align}
&\frac{d}{ds} \left[ \Delta \dot z^i  - \widetilde\phi^i_{\,\, a'}(\beta, s)\dot\beta^{a'} - \left. \frac{\partial\psi^i}{\partial\beta^{a'}} \right|_{\chi_0(s)} \! \Delta\beta^{a'} - \left. \frac{\partial\psi^i}{\partial\dot\beta^{a'}} \right|_{\chi(s)} \! \dot\beta^{a'} \right.
\nonumber\\
& \qquad\qquad \left. - \frac{1}{2} \left. \frac{\partial^2\psi^i}{\partial\dot\beta^{a'}\partial\dot\beta^{b'}} \right|_{\chi(s)} \! \dot\beta^{a'} \dot\beta^{b'} - \widetilde\psi^i_{\,\, a'}(\beta, s)\ddot\beta^{a'} \right]
\nonumber\\
&=   \left. \frac{\partial\phi^i}{\partial\beta^{a'}} \right|_{\chi_0(s)} \Delta\beta^{a'} + \left. \frac{\partial\phi^i}{\partial\dot\beta^{a'}} \right|_{\chi(s)} \dot\beta^{a'} 
\nonumber\\
&+ \frac{1}{2}\left. \frac{\partial^2\phi^i}{\partial\dot\beta^{a'}\partial\dot\beta^{b'}} \right|_{\chi(s)} \dot\beta^{a'}\dot\beta^{b'} - \left( \frac{\partial \widetilde\phi^i_{\,\, a'}}{\partial \beta^{b'}} \dot\beta^{b'} + \frac{\partial \widetilde\phi^i_{\,\, a'}}{\partial s} \right) \dot\beta^{a'} \,\, ,
\label{DeltazapproxsW}
\end{align}
where 
\begin{equation}
\chi(s) = (\beta, 0, s)
\end{equation}
and $\chi_0(s) = (\beta_0, 0, s)$. For small $s$ (of the order of the cycle duration), we typically have that $\Delta\beta \sim \dot\beta s$ and $\ddot\beta \sim \dot\beta/s$. Therefore, the order of magnitude of each term in the equation above, after the time integration, can be estimated as
\begin{align}
&\Delta \dot z^i  - \widetilde\phi^i_{\,\, a'}\dot\beta^{a'} - \frac{\partial\psi^i}{\partial\beta} \dot\beta s - \frac{\partial\psi^i}{\partial\dot\beta^{a'}}  \dot\beta^{a'} 
\nonumber\\
&- \frac{1}{2} \frac{\partial^2\psi^i}{\partial\dot\beta^{a'} \partial\dot\beta^{b'}}  \dot\beta^{a'} \dot\beta^{b'} - \widetilde\psi^i_{\,\, a'}\dot\beta^{a'} s^{-1} \, \sim
\nonumber\\
&\sim \, \frac{\partial\phi^i}{\partial\beta^{a'}}  \dot\beta^{a'}s^2 + \frac{\partial\phi^i}{\partial\dot\beta^{a'}}  \dot\beta^{a'}s + \frac{1}{2} \frac{\partial^2\phi^i}{\partial\dot\beta^{a'}\partial\dot\beta^{b'}}  \dot\beta^{a'} \dot\beta^{b'} s
\nonumber\\
& - \left( \frac{\partial \widetilde\phi^i_{\,\, a'}}{\partial \beta^{b'}} \dot\beta^{b'} + \frac{\partial \widetilde\phi^i_{\,\, a'}}{\partial s} \right) \dot\beta^{a'} s \,\, .
\label{Deltazapproxs}
\end{align}
Thus, assuming that any terms that go to zero when $s$ is taken to zero (while keeping all other parameters fixed) can be neglected, we obtain the following first order differential equation for $\Delta z$,
\begin{align}
&\Delta \dot z^i = \widetilde\psi^i_{\,\, a'}(\beta, s)\ddot\beta^{a'} \,+
\nonumber\\
&\left( \widetilde\phi^i_{\,\, a'}(\beta, s) +\frac{\partial\psi^i}{\partial\dot\beta^{a'}} (\beta, 0, s) + \frac{1}{2} \frac{\partial^2\psi^i}{\partial\dot\beta^{a'} \partial\dot\beta^{b'}} (\beta, 0, s) \, \dot\beta^{b'} \right) \dot\beta^{a'} .
\end{align}
Under the same $s \rightarrow 0$ condition, we may consider a Taylor expansion in $s$, so that, also assuming that the terms proportional to $s$ can be neglected, we get
\begin{align}
&\Delta \dot z^i =   \widetilde\psi^i_{\,\, a'}(\beta, 0)\ddot\beta^{a'} \,+
\nonumber\\
&\left( \widetilde\phi^i_{\,\, a'}(\beta, 0) +\frac{\partial\psi^i}{\partial\dot\beta^{a'}} (\beta, 0, 0) + \frac{1}{2} \frac{\partial^2\psi^i}{\partial\dot\beta^{a'} \partial\dot\beta^{b'}} (\beta, 0, 0) \, \dot\beta^{b'} \right) \dot\beta^{a'} ,
\end{align}
that has the form
\begin{equation}\label{Deltadotzeq}
\Delta \dot z^i =   \lambda^i_{\,\, a'}(\beta, \dot\beta) \dot\beta^{a'} + \widetilde\lambda^i_{\,\, a'}(\beta)\ddot\beta^{a'}  \,\, ,
\end{equation}
where
\begin{align}
\lambda^i_{\,\, a'}(\beta, \dot\beta) &= \widetilde\phi^i_{\,\, a'}(\beta, 0) +\frac{\partial\psi^i}{\partial\dot\beta^{a'}} (\beta, 0, 0) \,+
\nonumber\\
& \qquad \qquad + \frac{1}{2} \frac{\partial^2\psi^i}{\partial\dot\beta^{a'} \partial\dot\beta^{b'}} (\beta, 0, 0) \, \dot\beta^{b'}
\nonumber\\
\widetilde\lambda^i_{\,\, a'}(\beta) &= \widetilde\psi^i_{\,\, a'}(\beta, 0) \, .
\label{lambdaiadef}
\end{align}
Note that $\widetilde\phi$ is essentially related to tension contributions to the force, $\partial\psi/\partial\dot\beta$ is related to linear momentum contributions to the torque or spin-force coupling, $\partial^2\psi/\partial\dot\beta^2$ is related to tension contributions to the torque or spin-force coupling, and $\widetilde\psi$ to tension contributions to the torque or spin-force coupling.

This equation is not of the geometric-phase kind for it differs from equation (\ref{geophase}). Nevertheless, it still has relatively simple mathematical properties. The resulting motion depends on the image of the path traversed in the tangent bundle of the internal space. In order to be precise, let the {\sl extended state space} $\widetilde {\ca P}$ be defined as the vector bundle $(T\ca I \times \ca Z, T\ca I, \rho_1)$, where $\rho_1$ is the Cartesian projection of $T\ca I \times \ca Z$ on the first component, $T\ca I$. Given a curve $\gamma: \mathbb R \rightarrow \ca I$ in the configuration space, consider its canonical lift $ \widetilde\gamma : \mathbb R \rightarrow T\ca I$ to the tangent bundle (which corresponds to the field of tangent vectors along $\gamma$). For a coordinate system $\{\beta^{a'}\}$ in $\ca I$, take the natural coordinate system $\{ \beta^{a'}, \bar\beta^{b'} \}$ associated with $T\ca I$. If the coordinates of $\gamma$ are $\gamma^{a'}(s)$, so that $(\widetilde\gamma)^{a'} = \dot\gamma^{a'} = d\gamma^{a'}/ds$, then the coordinates of the canonical lift are given by $(\gamma^{1}(s), \ldots \gamma^{d}(s) ,  \dot\gamma^{1}, \ldots \dot\gamma^{d}(s))$. Consider a generalization of equation (\ref{Deltadotzeq}) given by
\begin{equation}\label{Deltadotzeq2a}
\Delta \dot \zeta^i =   \lambda^i_{\,\, a'}(\beta, \dot\beta, \zeta) \dot\beta^{a'} + \widetilde\lambda^i_{\,\, a'}(\beta, \dot\beta, \zeta)\ddot\beta^{a'}  \,\, ,
\end{equation}
which clearly reduces to (\ref{Deltadotzeq}) when $\lambda$ and $\widetilde\lambda$ are independent of $\zeta$. Based on this equation we define\footnote{Consider that the notation is analogous to the one employed before, except that now vectors in the base space are represented by $(v^{a'}, \bar v^{a'})$ instead of simply $(v^{a'})$.} the $\widetilde\Lambda$-lift of a vector $\widetilde v$ in $T\ca I$ to a vector $\widetilde V$ in $\widetilde {\ca P}$ as
\begin{equation}
\widetilde V = \widetilde\Lambda \widetilde v = \left( v^{a'}, \bar v^{a'}; \lambda^i_{\,\, a'}(\beta, \bar\beta, \zeta) v^{a'} + \widetilde\lambda^i_{\,\, a'}(\beta, \bar\beta, \zeta){\bar v}^{a'} \right) \,\, ,
\end{equation}
where $\widetilde v = (v^{a'}, \bar v^{a'})$ is an arbitrary vector at $T_{(\beta, \bar\beta)}T\ca I$ and $\widetilde V$ is a vector at $T_{(\beta, \bar\beta; \zeta)}\ca P$. Thus, a curve $\gamma$ in $\ca I$ can be canonically lifted to a curve $\widetilde\gamma$ in $T\ca I$, which then can be lifted to a curve $\widetilde\Gamma$ in $\widetilde{\ca P}$ defined analogously to the lift from $\ca I$ to $\ca P$ discussed before. It is therefore clear that the curve $\widetilde\Gamma$ represents a solution of equation (\ref{Deltadotzeq2a}). It is also clear that the deviations of the dynamical quantities $\Delta\zeta^i$ will be functions of the image of $\widetilde\Gamma$, and thence functions of the image of $\widetilde\gamma$. However, we see that different parameterizations for the curve $\gamma$ in the configuration space lift to different curves $\widetilde\gamma$ in the tangent space. Therefore, the deviations of the dynamical quantities will not be functions of the image of the curve in the configuration space, i.e.,
\begin{equation}
\Delta\zeta = \widetilde F(\mathrm{Im}(\widetilde\Gamma)) = \widetilde f(\mathrm{Im}(\widetilde\gamma)) \ne f(\mathrm{Im}(\gamma)) \,\, ,
\end{equation}  
where $\widetilde F$, $\widetilde f$ and $f$ are suitable functions. Thus, the overall motion of a body described by equation (\ref{Deltadotzeq}) does not depend solely on the sequence of geometrical shapes assumed by the body, but also on how fast the internal motions are performed.

This suggests that swimming (at low Reynolds number) in spacetime can only occur in very particular scenarios in which the interaction between the multipole moments and the curvature is such that $\widetilde\lambda$ is negligible and $\lambda$ does not depend on $\dot\beta$. This happens if there is no tension contributions to the torque and spin-force couple. More precisely, if the following vanish, or can be neglected in some regime,
\begin{align}
\widetilde\lambda^i_{\,\, a'} &= \frac{1}{M_0^2} S_0^{ij}  \frac{\partial F_j}{\partial \ddot\beta^{a'}} + \frac{1}{M_0}  \frac{\partial Q^{i0}}{\partial\ddot\beta^{a'}} = 0
\nonumber\\
\frac{\partial^2\psi^i}{\partial\dot\beta^{a'} \partial\dot\beta^{b'}} &=  \frac{1}{M_0^2} S_0^{ij}  \frac{\partial^2 F_j}{\partial\dot\beta^{a'} \partial\dot\beta^{b'}} + \frac{1}{M_0}  \frac{\partial^2 Q^{i0}}{\partial\dot\beta^{a'} \partial\dot\beta^{b'}} = 0 \,\, ,
\end{align}
where the right-hand sides are evaluated at $\dot\beta = 0$, $\ddot\beta = 0$ and $s=0$.

Consider first that we are only interested in observing swimming at order $(l/r)^2$, then we have to ask for the condition
\begin{equation}
\frac{\partial\bar J^{\nu\rho\sigma[i}}{\partial\ddot\beta^{a'}} {R_{\nu\rho\sigma}}^{0]} = \frac{\partial^2\bar J^{\nu\rho\sigma[i}}{\partial\dot\beta^{a'} \partial\dot\beta^{b'}} {R_{\nu\rho\sigma}}^{0]} = 0 \,\, ,
\end{equation}
and, since in the Newtonian regime the tensions only contribute to components of the multipole moments with the form $J^{ijkl}$ (i.e., with all space-like indices), then this condition reduces to
\begin{equation}
\frac{\partial\bar J^{ijkl}}{\partial\ddot\beta^{a'}} {R_{ijk}}^{0} = \frac{\partial\bar J^{ijkl}}{\partial\dot\beta^{a'} \partial\dot\beta^{b'}} {R_{ijk}}^{0} =  0 \,\, ,
\end{equation}
which can be satisfied for all bodies if the spacetime is such that
\begin{equation}
{R_{ijk}}^{0} = 0 \,\, ,
\end{equation}
recalling that the $0$-index is defined by the initial dynamical velocity (and the spatial indices are orthogonal to it). If ${R_{ijk}}^{0} \ne 0$, then only bodies with certain geometric shapes or performing specific internal motions could satisfy the condition above.

If one wants to observe swimming at $(l/r)^3$ order, then one must ask for the additional conditions
\begin{align}
&2M_0 \frac{\partial \bar J^{\nu\rho\sigma\lambda[i}}{\partial \ddot\beta^{a'}} \left( 3 \nabla_\nu {R_{\rho\sigma\lambda}}^{0]} + 2 \nabla_\lambda {R_{\nu\rho\sigma}}^{0]} \right) \,+
\nonumber\\
& \qquad \qquad + S_0^{ij} \frac{\partial \bar J^{\nu\rho\sigma\lambda}}{\partial \ddot\beta^{a'}} \nabla_j R_{\nu\rho\sigma\lambda}= 0
\nonumber\\
&2M_0 \frac{\partial^2 \bar J^{\nu\rho\sigma\lambda[i}}{\partial\dot\beta^{a'} \partial\dot\beta^{b'}} \left( 3 \nabla_\nu {R_{\rho\sigma\lambda}}^{0]} + 2 \nabla_\lambda {R_{\nu\rho\sigma}}^{0]} \right) \,+
\nonumber\\
& \qquad \qquad + S_0^{ij} \frac{\partial^2 \bar J^{\nu\rho\sigma\lambda}}{\partial\dot\beta^{a'} \partial\dot\beta^{b'}} \nabla_j R_{\nu\rho\sigma\lambda} = 0 \, ,
\end{align}
which, in the Newtonian regime for internal motions, reduce to
\begin{align}
&M_0 \frac{\partial \bar J^{klmri}}{\partial \ddot\beta^{a'}} \left( 3 \nabla_k {R_{lmr}}^{0} +  2\nabla_r {R_{klm}}^{0} \right) \,+
\nonumber\\
& \qquad \qquad+ S_0^{ij} \frac{\partial \bar J^{klmr}}{\partial \ddot\beta^{a'}} \nabla_j R_{klmr} = 0
\nonumber\\
&M_0 \frac{\partial^2 \bar J^{klmri}}{\partial\dot\beta^{a'} \partial\dot\beta^{b'}} \left( 3 \nabla_k {R_{lmr}}^{0} +  2\nabla_r {R_{klm}}^{0} \right) \,+
\nonumber\\
& \qquad \qquad+ S_0^{ij} \frac{\partial^2 \bar J^{klmr}}{\partial\dot\beta^{a'} \partial\dot\beta^{b'}} \nabla_j R_{klmr}= 0 \, ,
\end{align}
which can be satisfied for all bodies if
\begin{equation}
\nabla_l {R_{ijk}}^{\mu} = 0 \,\, ,
\end{equation}
otherwise only particular shapes and internal motions could satisfy them. Note that if there were no spin involved, then we could ask simply for $\nabla_l {R_{ijk}}^{0} = 0$.

Under these conditions, the equation for the position deviation is given by
\begin{equation}\label{Deltadotzeq2}
\Delta \dot z^i =   \lambda^i_{\,\, a'}(\beta) \dot\beta^{a'} \,\, ,
\end{equation}
manifesting the geometric-phase character of the dynamics. We define the {\sl differential displacement} as the 1-form in the configuration space $\ca I$ given by
\begin{equation}\label{diffadddisp}
\delta z^i =   \lambda^i_{\,\, a'}(\beta) d\beta^{a'} \,\, ,
\end{equation}
where $\lambda$ is defined in (\ref{lambdaiadef}), but can be written explicitly as
\begin{equation}\label{lambdaiaexplicit}
\lambda^i_{a'}(\beta) = \frac{1}{M_0} \frac{\partial F^i}{\partial \ddot\beta^{a'}} + \frac{1}{M_0^2} S_0^{ij} \frac{\partial F_j}{\partial \dot\beta^{a'}} + \frac{1}{M_0} \frac{\partial Q^{i0}}{\partial \dot\beta^{a'}} \,\, ,
\end{equation}
in which the right-hand side is being evaluated at $(\beta, \dot\beta, \ddot\beta, s) = (\beta, 0, 0, s_0)$. It is then clear that linear momenta and tensions are ``thrusting'' the swimming body.

When the body goes along a curve $\gamma$ in configuration space, the total displacement $\Delta z^i$ is given by
\begin{equation}\label{adddisp}
\Delta z^i = \int_\gamma \delta z^i = \int_\gamma  \lambda^i_{\,\, a'}(\beta) d\beta^{a'} \,\, .
\end{equation}
For a cyclic sequence of internal motions, we get
\begin{equation}\label{adddispcycle}
\Delta z^i_{cycle} =   \int_{\ca D} d\left( \lambda^i_{\,\, a'}(\beta) d\beta^{a'} \right) = \int_{\ca D}  \frac{\partial \lambda^i_{\,\, b'}}{\partial \beta^{a'}} \, d\beta^{a'}\!\! \wedge d\beta^{b'}  \,\, ,
\end{equation}
where $\ca D$ is a 2-dimensional surface in $\ca I$ enclosed by $\gamma$. If the cycle is a small parallelogram defined by the infinitesimal vectors $u^{a'}$ and $v^{a'}$ in $\ca I$, then formula (\ref{SwimmingFormula}) can be used,
\begin{equation}\label{SwimmingFormula2}
\Delta z^i_{cycle} \approx 2 \partial_{a'} \lambda^i_{\,\, b'} u^{[{a'}} v^{b']} \,\, ,
\end{equation}
which is the final formula for swimming (at low Reynolds number) in curved spacetimes.

\section{Comment on the conditions}
\label{sec:Conditions}

One should keep in mind that the approximations in (\ref{Deltazapproxs}) are not obviously valid in general, and it must be verified that they are indeed satisfactory in the particular case of analysis. In order to show how certain terms dropped in (\ref{DeltazapproxsW}) could often be problematic, let us consider a hypothetical case in which one of the configuration coordinates is $l$, representing the overall spatial size of the body. Assume that the mass distribution of the body contributes to $\psi$ according to
\begin{equation}\label{bodyexmass}
\psi_{mass} = A \frac{l^2}{r^2} + B \frac{l^3}{r^3} + \cdots \,\, ,
\end{equation}
where $A$ and $B$ are functions of the other configuration coordinates, and they are assumed to have the same order of magnitude $\sim \! kM$. The triple dots include mass-like terms that are of higher order in $l/r$ or that do not depend on $l$. Similarly, let the contribution from the linear momentum distribution to $\psi$ be
\begin{equation}\label{bodyexmom}
\psi_{momentum} = A' \frac{l^2}{r^2} \dot l + B' \frac{l^3}{r^3} \dot l + \cdots \,\, ,
\end{equation}
where $A'$ and $B'$ are other functions of the other configuration parameters, also of order of magnitude $\sim \! kM$. The triple dots here denote terms of higher order or not involving $\dot l$. The total $\psi$ is given by the sum of these two contributions
\begin{equation}
\psi = \psi_{mass} + \psi_{momentum}  \,\, .
\end{equation}
Observe that in the approximations (\ref{Deltazapproxs}), for very short integration times $s$, the term $(\partial\psi/\partial\beta)_{\chi_0} \Delta\beta$ was supposedly negligible compared to $(\partial\psi/\partial\dot\beta)_{\chi} \dot\beta$. In this model they become
\begin{align}
&\left. \frac{\partial\psi}{\partial\beta} \right|_{\chi_0} \Delta\beta = 2A \frac{l_0}{r^2} \Delta l + 3B \frac{l_0^2}{r^3} \Delta l + \cdots
\nonumber\\
&\left. \frac{\partial\psi}{\partial\dot\beta} \right|_{\chi} \dot\beta = A' \frac{l^2}{r^2} \dot l + B' \frac{l^3}{r^3} \dot l + \cdots \,\, ,
\end{align}
where $l = l_0 + \Delta l$. Hence, to the zeroth order of approximation in $l \approx l_0$, $(\partial\psi/\partial\beta)_{\chi_0} \Delta\beta$ differs from $(\partial\psi/\partial\dot\beta)_{\chi} \dot\beta$ by a factor $\Delta l/l \dot l \sim s/l$. For small times $s$, the first quantity is indeed negligible compared to the second; in general, we have only to ask for the condition
\begin{equation}
s \ll l \, ,
\end{equation}
which means that the time of the cycle should be much smaller than the spatial dimensions of the body. This is a case where the body cannot be centrally controlled, in ``real time'', for that would require faster-than-light signals propagating across its extension $-$ thus, that ingeniously-designed device proposed by Wisdom, coupled along the body, seems to be pertinent here (see beginning of section \ref{sec:multipole}). Note, however, that this condition only ensures that $(\partial\psi/\partial\beta)_{\chi_0} \Delta\beta$ is negligible compared to the lowest order approximation of $(\partial\psi/\partial\dot\beta)_{\chi} \dot\beta$. Consider, for instance, the first order expansion of $(\partial\psi/\partial\dot\beta)_{\chi} \dot\beta$ in $l$, around $l_0$,
\begin{align}
&\left. \frac{\partial\psi}{\partial\dot\beta} \right|_{\chi} \dot\beta \approx \left. \frac{\partial\psi}{\partial\dot\beta} \right|_{\chi_0} \dot\beta + \left. \frac{\partial}{\partial\beta} \left( \left. \frac{\partial\psi}{\partial\dot\beta} \right|_{\chi} \right) \right|_{\chi_0}\dot\beta \Delta\beta  =
\nonumber\\
& A' \frac{l_0^2}{r^2} \dot l + B' \frac{l_0^3}{r^3} \dot l + 2A' \frac{l_0}{r^2}\dot l \Delta l  + 3B' \frac{l_0^2}{r^3}\dot l \Delta l  + \cdots \,\, .
\end{align}
Although $(\partial\psi/\partial\beta)_{\chi_0} \Delta\beta$ is much smaller than the first two terms on the right-hand side of the expression above by a small factor $s/l$, the next two terms are at least of comparable magnitude since $\dot l \sim \Delta l/s \lesssim 1$, so that the motion does not exceed the speed of light. Therefore, even if no other term in (\ref{DeltazapproxsW}) were problematic, the resulting motion would be of the geometric-phase kind only up to first order in $\Delta\beta$. In particular, the swimming formula (\ref{SwimmingFormula2}) for closed cycles, being of order $\Delta\beta^2$, would become unreliable -- unless, of course, $(\partial\psi/\partial\beta)_{\chi_0}$ vanishes. At order $(l/r)^2$, $(\partial\psi/\partial\beta)_{\chi_0}$ is {\sl zero} if 
\begin{equation}
\frac{\partial\bar J^{\nu\rho\sigma[i}}{\partial\beta^{a'}} {R_{\nu\rho\sigma}}^{0]} = 0 \,\, ,
\end{equation}
which is evaluated at $\beta = \beta_0$ and $\dot\beta = 0$. Since only the mass distribution contributes to this, we may require simply
\begin{equation}
\frac{\partial\bar J^{j0 k0}}{\partial\beta^{a'}} {R_{j0 k}}^{i} = 0 \,\, ,
\end{equation}
and, at order $(l/r)^3$, we must also ask for
\begin{align}
&2M_0 \frac{\partial \bar J^{\nu\rho\sigma\lambda[i}}{\partial \beta^{a'}} \left( 3 \nabla_\nu {R_{\rho\sigma\lambda}}^{0]} + 2 \nabla_\lambda {R_{\nu\rho\sigma}}^{0]} \right) \,+
\nonumber\\
& \qquad \qquad + S_0^{ij} \frac{\partial \bar J^{\nu\rho\sigma\lambda}}{\partial \beta^{a'}} \nabla_j R_{\nu\rho\sigma\lambda} = 0 \, ,
\end{align}
which reduces to
\begin{align}
&S_0^{ij} \frac{\partial \bar J^{k0l0}}{\partial \beta^{a'}} \nabla_j R_{k0l0}+ 2M_0 \frac{\partial \bar J^{m0j0i}}{\partial \beta^{a'}} \nabla_0 {R_{m0j}}^{i} \, - 
\nonumber\\
&-\, M_0 \frac{\partial \bar J^{mj0k0}}{\partial \beta^{a'}} \left( 3 \nabla_m {R_{j0k}}^{i} + 2 \nabla_k {R_{m(0j)}}^{i} \right) = 0 \,\, .
\end{align}
In order to have these conditions automatically fulfilled regardless of the body shape and internal motions, the curvature tensor and its first derivative have to satisfy $R_{ijk0} = 0$ (which is the only condition imposed at $(l/r)^2$ order), $\nabla_\mu R_{ijk0} = 0$ and, if there is spin, $\nabla_j R_{k0l0} = 0$. 

In the general case, it seems that most of the terms in the right-hand side of (\ref{DeltazapproxsW}) can also impair the validity of formula (\ref{SwimmingFormula2}) at $(l/r)^3$ order. Nevertheless, even if that happens, the overall motion could still be swimming-like up to first order in the amplitudes $\Delta\beta$ of the internal motions. Thus, formula (\ref{adddisp}) would still be reliable up to first order in the amplitudes, i.e.,
\begin{equation}
\Delta z^i \approx   \lambda^i_{\,\, a'}(\beta_0) \Delta\beta^{a'}
\end{equation}
even at $(l/r)^3$ order.

Let us briefly consider these other terms. The term $(\partial\phi^i/\partial\beta^{a'})_{\chi_0} \Delta\beta^{a'}$ cannot cause problems since it only affects the swimming formulas at order $\Delta\beta^3$. The term $(\partial \widetilde\phi^i_{\,\, a'}/\partial s)\dot\beta^{a'}$ could only cause trouble if the multipole moments were explicitly dependent on time, as can be seen by
\begin{align}
&\frac{\partial \widetilde\phi^i_{\,\, a'}}{\partial s}\dot\beta^{a'} = \frac{1}{6M_0} \left[ \frac{\partial}{\partial s}\left( \frac{\partial {\bar J}^{jklm}}{\partial \ddot\beta^{a'}} \right) \nabla^i R_{jklm} + \right.
\nonumber\\
&\left. + \frac{\partial {\bar J}^{jklm}}{\partial \ddot\beta^{a'}} \frac{\partial }{\partial s} \left( \nabla^i R_{jklm} \right) \right] \dot\beta^{a'} \,\, ,
\end{align}
so that the first term above may affect the equations at the order $k/r^3$ while the second can be discarded for it is proportional to a second derivative of the curvature tensor (supposedly of order $k/r^4$). If tension-related moments were explicitly time-dependent, then we would have to require
\begin{equation}
\frac{\partial}{\partial s}\left( \frac{\partial {\bar J}^{jklm}}{\partial \ddot\beta^{a'}} \right) \nabla^i R_{jklm} = 0 \,\, ,
\end{equation}
which is satisfied for arbitrary bodies only if $\nabla_i R_{jklm} = 0$.

The quantity $(\partial\phi^i/\partial\dot\beta^{a'})_{\chi} \dot\beta^{a'}$ tends to be non-vanishing in most cases. The condition for it to vanish is basically that there should be no linear momentum contribution to the quadrupole-order force, which translates as
\begin{equation}
\frac{\partial {\bar J}^{\mu\nu\rho\sigma}}{\partial\dot\beta^{a'}} \nabla^i R_{\mu\nu\rho\sigma} = 0 \,\, ,
\end{equation}
that reduces to
\begin{equation}
\frac{\partial {\bar J}^{0jkl}}{\partial\dot\beta^{a'}} \nabla^i R_{0jkl} = 0 \,\, ,
\end{equation}
which can be generally satisfied if $\nabla_i R_{0jkl} = 0$. Note that it does not cause problems at order $(l/r)^2$.

The quantities $(1/2) (\partial^2\phi^i/\partial\dot\beta^{a'}\partial\dot\beta^{b'})_\chi \dot\beta^{a'}\dot\beta^{b'}$ and $(\partial \widetilde\phi^i_{\,\, a'}/\partial \beta^{b'}) \dot\beta^{a'}\dot\beta^{b'}$ are associated with the tension contributions to the force. They vanish if
\begin{equation}
\left( \frac{1}{2} \frac{\partial^2 \bar J^{\nu\rho\sigma\lambda}}{\partial\dot\beta^{a'}\partial\dot\beta^{b'}} - \frac{\partial^2 \bar J^{\nu\rho\sigma\lambda}}{\partial\beta^{a'}\partial\ddot\beta^{b'}} \right)  \nabla^i R_{\nu\rho\sigma\lambda} = 0 \,\, ,
\end{equation}
which reduces to
\begin{equation}
\left( \frac{1}{2} \frac{\partial^2\bar J^{jklm}}{\partial\dot\beta^{a'}\partial\dot\beta^{b'}} - \frac{\partial^2\bar J^{jklm}}{\partial\beta^{a'}\partial\ddot\beta^{b'}} \right)  \nabla^i R_{jklm} = 0 \,\, .
\end{equation}
Thus, a general body can satisfy this if $\nabla_i R_{jklm} = 0$. Again, no problem is posed at $(l/r)^2$ order.

Due to the many different conditions that we have obtained, it is appropriate to summarize them in a more organized manner. Note that we are mostly interested in the conditions that allow a general body to swim, despite of its geometrical shape or internal motions. So, we have found that swimming can be fully observed, up to order $(l/r)^2$, if only $R_{ijk0} = 0$. When investigating the dynamics at order $(l/r)^3$, however, it is possible that formula (\ref{adddisp}) could be correct only up to first order in the amplitudes $\Delta\beta$, so that formula (\ref{SwimmingFormula2}) would be unreliable $-$ in this case, we shall say that the body can {\sl quasi-swim}. When there is no spin, $\nabla_l R_{ijk0} = 0$ is enough to ensure that the body can quasi-swim, while $\nabla_0 R_{ijk0} = \nabla_l R_{ijkm} = 0$ is further required for a legitimate swimming. When there is spin, $\nabla_l R_{ijk0} = \nabla_l R_{ijkm} = 0$ is required for {\sl quasi-swimming}, while $\nabla_j R_{k0l0} = 0$ is additionally demanded for legitimate swimming (i.e., only the components $\nabla_0 R_{i0j0}$ could be non-vanishing). Recall that these components refer to the coordinate system associated with the initial conditions; covariantly, the condition $R_{ijk0} = 0$, for instance, would be translated as ``the restriction of $R(u, \cdot, \cdot, \cdot)$ to the vector subspace orthogonal to $u$ at $z(s_0)$ must vanish''.

We reinforce that all these conditions refer specifically to a body as introduced in this section  -- namely, one described by multipole moments as given in (\ref{bodyexmass}) and (\ref{bodyexmom}). In any particular case one must carefully go through similar steps to verify if the approximations in (\ref{Deltazapproxs}) are indeed reasonable. Still, this model seems to encompass a handful of interesting scenarios such as, to name a few, a bipod on a sphere (App.~\ref{sec:bipod}) and a tripod in Schwarzschild \cite{Rodrigo} or FLRW spacetime (App.~\ref{sec:FLRW}).

Finally, let us make a brief comment on the core assumption that the body is light and presumably does not affect the spacetime geometry. One may worry that in the limit of vary fast cycles the emission of gravitational waves could become relevant. To roughly estimate the effect of these gravitational waves, recall that in Minkowski spacetime, perturbatively, the power radiated away is given by
\begin{equation}
\mathcal P_{GW} = \frac{16}{45} \sum_{ij} \left( \frac{d^3 K^{ij}}{dt^3} \right)^2
\end{equation}
where $K^{ij}$ is the trace-free part of $J^{0i0j}$. 
In the assumptions of Sec.~\ref{octupole}, we would expect
\begin{equation}
\mathcal P_{GW} \sim \frac{M^2 \Delta l^4}{s^6}
\end{equation}
Further assuming that the reaction force could be estimated as $F_{GW} \sim \mathcal P_{GW}/(\Delta l/s)$, we would have
\begin{equation}
F_{GW} \sim \frac{M^2 \Delta l^3}{s^5}
\end{equation}
The important aspect to notice is that this force is proportional to $(\text{\sl mass})^2$, while the other forces and torques we have been considering are linear in the mass. Accordingly, if $F_{GW}$ is to be negligible in comparison with the lowest order term in \eqref{emSN3order}, we need
\begin{equation}
M \ll \frac{l s^5}{\Delta l^3 r^2}
\end{equation}
For a given mass, this imposes another constraint on how fast the cycles are allowed to be (in addition to $\Delta l \ll s$).

\section{Discussion}
\label{sec:Discussion}

The main purpose of this paper was to provide a rigorous and comprehensive treatment of the phenomenon of swimming (at low Reynolds number) in curved spacetime. We found that this effect can indeed take place, although only in very special circumstances where the spacetime abounds with symmetries and the initial conditions and body configurations are favorable. The framework here developed, deeply rooted in Dixon's multipole theory of dynamics, is not limited to the ``low Reynolds number'' regime, being suitable to describe the dynamics of any free small articulated body performing fast cyclic internal motions in general relativity. 

The choice of Dixon's formalism as the starting point was due to some of its important features. Namely, the fact that $(i)$ the fundamental conservation equation for matter in general relativity, $\nabla_a T^{ab} = 0$, is automatically implemented in the theory, $(ii)$ the multipole structure of the equations allow a controlled separation of the dynamical effects at different orders in $l/r$ and $(iii)$ it provides a reasonable definition for the worldline of the center of mass, which can be used to track the net motion of the body.

One important aspect of our approach was the special manner proposed to prescribe the quadrupole and octupole moments. More precisely, we have shown that, when solving the multipole equations up to $(l/r)^3$ order of precision, the relevant moments could be evaluated as if the background spacetime were flat, using the exponential map to establish the connection to the real (curved) spacetime. Moreover, the dynamics of the body could be described by the equations of special relativity, namely the energy-momentum tensor would satisfy the conservation equation $\partial_a T^{ab} = 0$, where $\partial$ is the derivative operator associated with any inertial frame of the flat spacetime. This means that, under the $(l/r)^3$ approximation, the problem of solving the dynamics in general relativity can be formally reduced to solving an analogous problem in special relativity. Obviously this amounts to a great simplification. 


Considerable attention was devoted to the special case of articulated bodies, with finite-dimensional internal configuration spaces, $\ca I$, performing a predetermined routine of internal motions, $\gamma : \mathbb R \rightarrow \ca I$. We found that, for small integration times and in the Newtonian regime for the internal movements, the effective motion can be described by a generalized concept of swimming. In this generalization, the deviation of dynamical quantities, with respect to the rigid body solution, are functions of the image of the (canonical) lift of $\gamma$ to the tangent space of the configuration space, $T\ca I$. Although this kind of motion is relatively simple, it does not display the defining feature of the traditional swimming at low Reynolds numbers. Indeed, the overall motion depends on the rapidity of the internal motions. It is interesting to recall that the Newtonian regime condition comes to ensure that the $\ddot\beta$-dependent part of the force does not depend on the ``velocities'' $\dot\beta$; therefore, this condition is not required at order $(l/r)^2$, possibly making this form of swimming a fairly general effect at this order of approximation (as long as certain terms can be neglected in the limit of fast cycles).

When there are no tension contributions to the torque, then it might be possible to swim or quasi-swim (at low Reynolds number) in spacetime. This is a very strong condition, not often satisfied. Moreover, this condition depends on initial values such as the initial dynamical velocity and spin. This can be understood by noticing that swimming is not possible in Newtonian gravitation because only the mass content of the body interacts with gravity, so that speeding up the cycles of internal motions implies that the gravitational impulse becomes smaller, influencing the resulting motion. In general relativity the linear momentum and tension densities also ``feel'' the gravitational field and, since they increase for faster cycles, they can compensate for the shorter cycle duration in such a way that the gravitational impulse becomes independent of its duration -- this is swimming at low Reynolds number. However, tension contributions to the torque tend to ``overcompensate'' for these shorter cycle times, in the sense that, at some point the impulse starts to increase for faster cycles and the resulting motion becomes dependent again on the rapidity of the internal motions.

Although swimming (at low Reynolds number) in curved spacetime is indeed a phenomenon of limited occurrence, there are some important spacetimes where swimming (or quasi-swimming) can take place. Namely, it happens in Friedmann-Lema\^itre-Robertson-Walker spacetimes, if the body starts at rest with respect to the family of isotropic observers. It is worth to point out that this scenario, where the spatial sections are homogeneous and isotropic, represents the closest analogue to the Newtonian swimmer in a non-turbulent viscous fluid. Indeed, the fluid mechanics theory of swimming at low Reynolds number, in which the effective motion is described by geometric phases associated with the sequence of shapes assumed by the body \cite{Wilczek}, is not expected to apply to a non-homogeneous (or non-isotropic) fluid, such as a fluid with non-constant density or viscosity.

In conclusion, the dynamics of small articulated bodies performing fast internal motions in curved spacetimes is much more complex than originally expected, as the effective motion can respond differently to different rapidities of the internal movements. Accordingly, we propose that the term {\sl swimming in curved spacetime} should not be restricted to this limited sense of ``low Reynolds number'' regime, in the same way that the term {\sl swimming} in the context of fluids is not. Finally, there are many interesting directions for future exploration in this subject. For example, how a spacetime swimmer could become more efficient (i.e., what is the regime, shape and internal motions required to achieve the maximum displacement per cycle)? Some recent work that have addressed questions of this nature include \cite{Harte,Harte2,harte2023local}.

\acknowledgments

I would like to thank George Matsas and Daniel Vanzella for helpful discussions, and also for comments on the first version of this paper. I also thank Ted Jacobson for valuable comments on the final version of this paper. I acknowledge full financial support from S\~ao Paulo Research Foundation (FAPESP) under Grant 2015/10373-4 during the time this work was produced, and full financial support from the University of Maryland, College Park under Grant NSF PHY-1708139 during the time this paper was written.

\appendix
\section{Definition of relevant bitensors}
\label{sec:bitensors}

In section (\ref{sec:multipole}) we presented formula (\ref{explicJ2}) giving the multipole moments in terms of the stress-energy tensor. In this appendix we provide the definitions for the bitensors appearing in that formula. 

We define the {\sl world-function} biscalar $\sigma$ by
\begin{equation}\label{sigma}
\sigma(x_1,x_2) = \frac{1}{2} (u_2 - u_1) \int^{u_2}_{u_1} g_{ab}v^a v^b du \,\, ,
\end{equation}
where $x(u)$ is an affine-parametrized geodesic linking  $x_1 = x(u_1)$ and $x_2 = x(u_2)$, and $v^a$ is the tangent vector field to it. Note that, in the hypothesis of normal neighborhood, there is a single geodesic linking each two points, making this bi-scalar well-defined.

We represent higher-rank {\sl bitensors}, based on the pair of points $(x,z)$, using a convention for the indices similar to the abstract index notation for tensors. The difference is that we reserve the first half of the Latin alphabet (``$a,b,c,\ldots$'') to denote slots associated with the point $x$ (i.e., the first entry) and the second half of the alphabet (``$k,l,m,\ldots$'') to denote slots associated with the point $z$ (i.e., the second entry). Components associated with $x$ will be denoted by letters in the first half of the Greek alphabet (``$\alpha, \beta,\gamma,\ldots$'') and those associated with $z$ by letters in the second half (``$\kappa,\lambda,\mu,\ldots$''). We can act with covariant derivatives on bitensors by keeping the non-involved point fixed (and also ignoring the tensor indices associated with that point). Whenever the argument of bitensors are omitted, it is implicit that they are $(x,z)$.

When covariant derivatives act on $\sigma$ we denote it as, for example, $\nabla_a \nabla_b \nabla_k \sigma = \sigma_{kba}$. In particular, we have $\sigma^a$ and $\sigma^k$. Also, when a second derivative acts on $\sigma^k$, now with respect to the point $x$, we get $\bi{\sigma}{k}{a}$. This bitensor can be seen as a (linear) mapping of vectors at $x$ to vectors at $z$, i.e., a map of the kind $T_x \mathcal{M}\rightarrow T_z\mathcal{M}$ (if $v^a \in T_x \mathcal M$, then $\bi{\sigma}{k}{a} v^a \in T_z \mathcal M$). In a region such that every point is in the normal neighborhood of the others, this is an invertible map, and we define the inverse map ${\widehat \sigma}^a_{\:\: k}$ by
\begin{equation}\label{biinv}
{\widehat \sigma}^a_{\:\: k} \sigma^k_{\:\:b} = \delta^a_{\:\:b} \,\, ,
\end{equation}
where $\delta^a_{\:\:b}$ is the identity tensor at $x$. We also define the {\sl transport bitensors} $K$ and $H$ by
\begin{align}
&K^a_{\:\:k} =- {\widehat \sigma}^a_{\:\: l} \sigma^l_{\:\:k}
\nonumber\\
&H^a_{\:\:k} =- {\widehat \sigma}^a_{\:\: k} \, ,
\label{KH}
\end{align}
which can be thought as (linear) mappings of vectors at $z$ to vectors at $x$ (i.e., maps of the kind $T_z \mathcal{M}\rightarrow T_x\mathcal{M}$). Finally, define the {\sl $n$-potential function}, for $n \ge 2$, as
\begin{equation}\label{npotential}
{\langle n \rangle}^{klqr}(x,z) = (n-1) \int_0^1 \sigma^{ka}\bi{\sigma}{(q}{\,\,a}\bi{\sigma}{r)}{\,b} \sigma^{lb} u^{n-2} du \, ,
\end{equation}
where the arguments of the $\sigma$'s are $(\overline x(u), z)$, with $\overline x(u)$ being a geodesic with affine parameter $u$ such that $\overline x(0) = z$ and $\overline x(1) = x$.

\section{A bipod on a sphere}
\label{sec:bipod}

In this appendix we analyze the problem of a bipod constrained to move on a 2-sphere. This problem was one of the major motivations behind Wisdom's realization that swimming (at low Reynolds number) in the curved spacetimes of general relativity could be possible. We first review the Newtonian results, and then we apply the covariant theory of swimming to the equivalent problem in general relativity, showing that the answers match. Hence, this serves as a check of consistency for our theory.

A bipod is set to move freely (i.e., without friction) on a 2-sphere of radius $r$, which will be covered with spherical coordinates $\{\phi, \psi\}$, where $\phi = [0, \pi]$ and $\psi = [0, 2\pi)$. The bipod is formed by a central mass $m_0$ placed (without loss of generality) at the point $(\pi/2, 0)$, connected by two straight (geodesic) massless rods of length $l$ to two other identical masses $m_1 = m_2$. The bipod is allowed to stretch and shrink its legs or open and close the angle $2\alpha$ between the legs at $m_0$; hence, the associated internal configuration space $\ca I$ is spanned by the coordinates $\{\alpha, \beta\}$, where $\beta = l/r$.

Let the axis of symmetry (bisectrix of the bipod) be initially aligned with the $\psi$ direction; thence, because of the symmetry of the bipod and the sphere, the effective motion of the bipod will necessarily be along the $\psi$ direction, and the bisectrix direction will not tilt. Therefore, the only meaningful dynamical variable is the $\psi$ coordinate of, say, the central mass. The (Newtonian) differential displacement\footnote{Recall that the differential displacement 1-form is defined from the equation satisfied by the trajectory deviation, $\Delta\psi(t) = \psi(t) - \psi_0(t)$, where $\psi_0(t)$ is the trajectory of the rigid bipod. Namely, if $\Delta\dot\psi = A\, \dot l + B\, \dot\alpha$, then $\delta\psi = A\, dl + B\, d\alpha$.} is given by
\begin{align}
&\delta \psi = \frac{2m_1 \sin\!\beta \cos\!\beta \sin\!\alpha}{m_0 + 2m_1 (\cos^2\!\beta + \sin^2\!\beta \cos^2\!\alpha)} \, d\alpha 
\nonumber\\
&- \frac{2m_1 \cos\!\alpha}{m_0 + 2m_1 (\cos^2\!\beta + \sin^2\!\beta \cos^2\!\alpha)} \, d\beta \, .
\label{Ndad2}
\end{align}
After a closed cycle of internal motions, the net additional displacement is
\begin{equation}
\Delta \psi = \int_{\mathcal D} \left[ - \frac{4 m_0 m_1 \sin\!\alpha \sin^2\!\beta}{(m_0 + 2m_1 (\cos^2\!\alpha + \cos^2\!\beta \sin^2\!\alpha))^2} \right] \, d\beta \wedge d\alpha \,\, .
\end{equation}
where $\ca D$ is the region in $\ca I$ enclosed by the cycle. Thus, considering the rectangular path $(\beta, \alpha) \rightarrow (\beta + \Delta\beta, \alpha) \rightarrow (\beta + \Delta\beta, \alpha+ \Delta\alpha) \rightarrow (\beta, \alpha + \Delta\alpha) \rightarrow (\beta, \alpha)$ in the configuration space, with $\Delta\beta$ and $\Delta\alpha$ small amplitudes, we can approximate
\begin{equation}
\Delta \psi \approx  - \frac{4 m_0 m_1 \sin\!\alpha \sin^2\!\beta}{(m_0 + 2m_1 (\cos^2\!\alpha + \cos^2\!\beta \sin^2\!\alpha))^2} \,   \Delta\beta \Delta\alpha \,\, .
\end{equation}
For a small bipod ($l/r \ll 1$), the Taylor expansion in $\beta$ yields
\begin{align}
&\Delta \psi \approx  - \frac{4 m_0 m_1 \sin\!\alpha}{(m_0 + 2m_1)^2} \, \beta^2  \Delta\beta \Delta\alpha +
\nonumber\\
& \frac{4 m_0 m_1 \left( m_0 + 12m_1 \cos^2\!\alpha - 10 m_1 \right) \si\alpha}{3(m_0 + 2m_1)^3} \, \beta^4 \Delta\beta \Delta\alpha + \cdots \,\, ,
\end{align}
so that the lowest order effect is at order $(l/r)^2$ and there is no correction at order $(l/r)^3$. Hence, we expect the result from the covariant theory of swimming to be
\begin{equation}
\Delta \psi =  - \frac{4 m_0 m_1 \sin\!\alpha}{(m_0 + 2m_1)^2} \, \beta^2  \Delta\beta \Delta\alpha \,.
\end{equation}
Fortunately, this is precisely what we will obtain.

Let us now look at this problem as the limiting case of a slowly moving bipod in the context of general relativity. The simplest thing that one can think of is to consider a spacetime with topology $\mathbb{R}\times S^2$ and metric $ds^2 = dt^2 - r^2 (d\phi^2 + \sin^2\!\phi \, d\psi^2 )$, where $t \in \mathbb R$, $\phi \in [0,\pi]$ and $\psi \in [0,2\pi)$. However, in order to avoid going through the necessary modifications in our formulas associated with the different spacetime dimension, we shall have the bipod constrained in a spatial 2-sphere {\sl inside} a 1+3 spacetime. 

In the Newtonian case, there was no issue in considering the bipod to be held on the sphere by normal {\sl forces of constraint} which could, for instance, be of contact nature: the Lagrangian formalism deals well with this. The Dixon's formalism, on the other hand, provides no simple way of addressing constraints like this. The solution to this situation is to realize that the swimming of the bipod on the sphere should not depend on the nature of the forces of constraint. Thus, we might search for a spacetime in which this force is of {\sl gravitational} nature, allowing us to directly apply Dixon's formalism (as the body would be essentially free in this case). The natural candidate for the spacetime is one with topology $\mathbb{R}\times S^3$ and metric
\begin{equation}\label{gRS3}
ds^2 = dt^2 - r^2 (d\theta^2 + \sin^2\!\theta \, d\phi^2 +\sin^2\!\theta \sin^2\!\phi \, d\psi^2 ) \,\, ,
\end{equation}
where $\theta \in [0,\pi]$, $\phi \in [0,\pi]$ and $\psi \in [0,2\pi)$. Note that the submanifold $\theta = \pi/2$ is clearly diffeomorphic to the spacetime $\mathbb{R}\times S^2$ mentioned above. Moreover, it has the convenient property that any geodesic tangent to it, at any point, lies entirely within it (such a submanifold is said to be {\sl totally geodesic}). If the bipod is initially placed on this submanifold, and all the internal motions are tangent to it, then the body will not ``take off'' and leave the submanifold as it evolves (which can be properly confirmed by analysis of the equations of motion).

The only non-vanishing components of the curvature tensor on $\mathbb{R}\times S^3$, in the orthonormal basis $e_\mu$ aligned with the coordinates $(t, \theta, \phi, \psi)$, restricted to the submanifold $\mathbb{R}\times S^2$, are given by
\begin{equation}
R_{1212} = R_{1313} = R_{2323} = \frac{1}{r^2} \, ,
\end{equation}
and all the derivatives of the curvature tensor vanish everywhere, $\nabla_a R_{bcde} = 0$. This means that the Dixon's equations become exact, with the force and torque given by
\begin{align}
&F^k = \frac{1}{2} R^k_{\:\: lqr} v^l S^{qr} 
\nonumber\\
&Q^{kl} = \frac{4}{3} J^{pqr[k} {R_{pqr}}^{l]} \,\, ,
\label{FQbipod}
\end{align}
Although the equations of motions are simple, the exact evaluation of the quadrupole moment can still be very complicated. Hence, let us assume that the body is sufficiently small compared to the radius (i.e., $l \ll r$), so that we can use the framework developed in this paper to obtain final results reliable up to $(l/r)^3$ order. 

According to formula (\ref{FQbipod}), the force is orthogonal to the dynamical velocity, implying that the proper mass $M$ is always conserved. From an analysis of the conserved quantities (see item $(\text{vi})$ in section \ref{sec:Dixon}) we find that, if the bipod is released from rest and with no spin, then the linear and angular momentum will not vary with time, that is,
\begin{align}
&p^k(s) = M (e_0)^k 
\nonumber\\
&S^{kl}(s) = 0
\end{align} 
for all $s$. Therefore the equation for the linear momentum becomes trivial, the one for the angular momentum yields
\begin{equation}
e_0^{[a}v^{b]} = - \frac{1}{2M} Q^{ab}
\end{equation}
and equation (\ref{uvmis}) is automatically satisfied. In components associated with the $e_\mu$ basis, we get
\begin{equation}\label{vibipod}
v^i = \frac{1}{M}Q^{i0} \, ,
\end{equation}
where $i$ denotes a ``spatial'' index. This can be integrated to yield the trajectory of the center of mass of the body. 

According to (\ref{FQbipod}), the non-vanishing components of the torque read
\begin{align}
&Q^{10} = \frac{4 v^0}{3r^2} \left( J^{2120} + J^{3130} \right)
\nonumber\\
&Q^{20} = \frac{4 v^0}{3r^2} \left( J^{1210} + J^{3230} \right)
\nonumber\\
&Q^{30} = \frac{4 v^0}{3r^2} \left( J^{1310} + J^{2320} \right) \,\, ,
\end{align} 
recalling that, in our approximations, $v^0 \approx 1$. Thus, only the momentum-related components of the quadrupole moment, $J^{iji0} = \int \!d^3\!x\, x^i x^{[i}g^{j]}$, are relevant. Since the bipod is contained to the 2-3 ``plane'' (i.e., the 2-sphere), then $J^{1j10}=0$; also, since there is no linear momentum in a direction orthogonal to the 2-sphere, then $J^{i1i0} = 0$. Therefore, the torque components simplify to
\begin{align}
&Q^{10} = 0
\nonumber\\
&Q^{20} = \frac{4 v^0}{3r^2} J^{3230} 
\nonumber\\
&Q^{30} = \frac{4 v^0}{3r^2}  J^{2320}  \,\, .
\label{bipodQi0}
\end{align} 
Furthermore, because of the inversion symmetry with respect to the bisectrix of the bipod, one should have $J^{3230} = J^{klmn}(e_3)_k (e_2)_l (e_3)_m (e_0)_n = J^{klmn}(e_3)_k (- e_2)_l (e_3)_m (e_0)_n = 0$, so that the only non-vanishing components of the torque are $Q^{30} = - Q^{03}$. This implies that the bipod moves along the great circle of the 2-sphere (i.e., the worldline of the center of mass lies in the $t$-$\psi$ submanifold), as expected. Parameterizing the worldline by the coordinate time $t$, we obtain from (\ref{vibipod}) the simple formula
\begin{equation}
\frac{d\psi}{dt} = \frac{4}{3} \frac{J^{2320}}{Mr^3} \,\, .
\end{equation}
Note that this result is correct up to $(l/r)^3$ order, and all the relativistic effects (e.g., those coming from very fast internal movements) must be encoded in $J^{2320}$. 

Let us now compute this quadrupole moment. We shall assume that the internal motions are sufficiently slow so that Newtonian mechanics apply. This is motivated by the general theory, which indicates that swimming (at low Reynolds number) is mainly limited to this regime. 

\begin{figure}[!h]
\centering
\includegraphics[scale = 0.6]{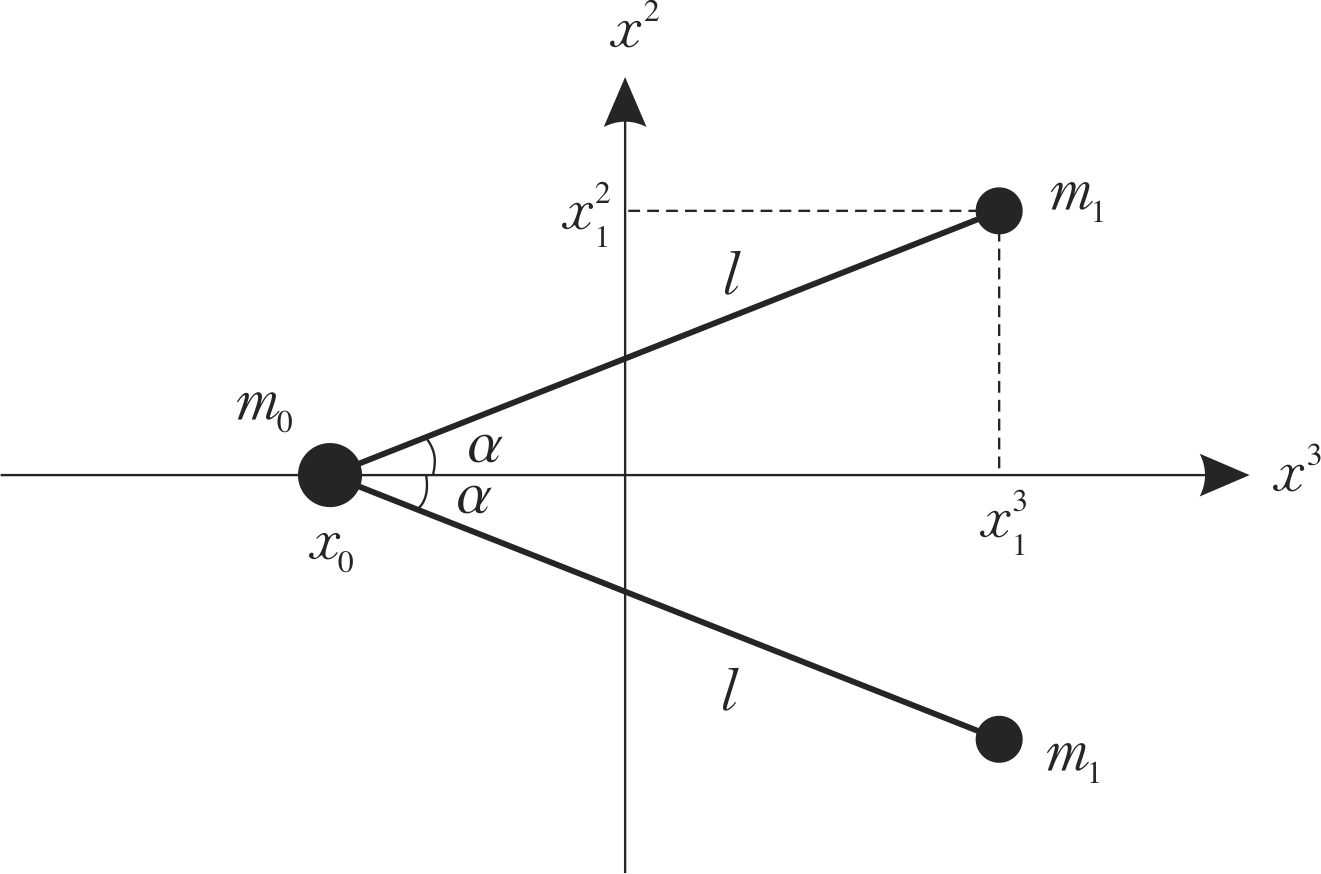}
\caption{The local correspondence to a flat spacetime puts the bipod on an Euclidean plane. The coordinate $x^2$ is aligned with the actual direction $e_2$, in the real spacetime, and $x^3$ is aligned with the actual direction $e_3$.}
\label{FIGswimbipodcoord}
\end{figure}

\noindent
The total mass is given by $M = m_0 + 2m_1$, and the momentum $g^j$ is
\begin{equation}
g^j(x) = \rho(x) \vartheta^j(x) \, ,
\end{equation}
where $\vartheta^j(x)$ is the 3-velocity of the element of mass at the point $x$.  To evaluate the component $J^{2320}$, we consider that the bipod is in an Euclidean 2-dimensional plane, covered with coordinates $\{x^2, x^3\}$, such that $x^2$ is in $e_2$ direction (i.e., $\phi$ direction) and $x^3$ is in $e_3$ direction (i.e., $\psi$ direction) [see figure (\ref{FIGswimbipodcoord})]. Let the coordinates of the central mass, $m_0$, be $(x_0^2, x_0^3) = (0, x_0)$, where $x_0$ is some number. The coordinates of the mass $m_1$, at an arbitrary configuration, are given by $(x_1^2, x_1^3) = (l \si\alpha, x_0 + l \co\alpha)$ and those of the mass $m_2$ are $(x_2^2, x_2^3) = (- l \si\alpha, x_0 + l \co\alpha)$. In order to have the center of mass at the origin of the coordinate system (which is essential to the definition of the quadrupole moment), we must set $x_0$ as
\begin{equation}
x_0 = - \frac{2 m_1}{m_0 + 2m_1} l \co\alpha \, .
\end{equation}
The velocities $\dot x_k^i$ can be easily computed and the density of momentum can be expressed as
\begin{equation}
g^i(x) = \sum_k m_k \delta(x - x_k) \dot x_k^i \,\, ,
\end{equation}
where $\delta$ is the {\sl Dirac delta function}. We find the component $J^{2320}$ to be
\begin{equation}\label{bipodJ2320}
J^{2320} = - \frac{m_0 m_1}{m_0 + 2m_1} l^3 \si\alpha \, \dot\alpha \, ,
\end{equation}
so that
\begin{equation}
\frac{d\psi}{dt} = - \frac{4}{3r^3} \frac{m_0 m_1}{(m_0 + 2m_1)^2} l^3 \si\alpha \, \frac{d\alpha}{dt} \,\, .
\end{equation}
When the bipod is rigid (i.e., $\dot l = 0$ and $\dot\alpha = 0$), the center of mass does not move, $\psi_0(t) = \psi(0)$. Thus the differential displacement $\delta\psi$ is
\begin{equation}\label{bipoddeltaphi}
\delta\psi = - \frac{4}{3r^3} \frac{m_0 m_1}{(m_0 + 2m_1)^2} l^3 \si\alpha \, d\alpha \,\, .
\end{equation}
Note that it differs from the corresponding 1-form (given in equation (\ref{Ndad2})) of the Newtonian approach, and this is because $\psi$ there denoted the position of the pivot mass while here it denotes the position of the center of mass. We expect, nonetheless, that after a complete cycle of internal motions the two distinct definitions for the net deviations should match -- after all, the distance between the pivot and the center of mass depends only on the shape of the bipod, which is the same before or after a cycle. After a complete cycle (of infinitesimal internal motions) the total deviation is
\begin{equation}
\Delta \psi  = - \frac{4 m_0 m_1}{(m_0 + 2m_1)^2} \beta^2 \si\alpha \, \Delta\beta \Delta\alpha \,\, ,
\end{equation}
which is precisely the correct result, up to $(l/r)^3$ order of approximation!

This result can be quickly re-obtained if we apply the swimming formula (\ref{SwimmingFormula2}). The differential displacement is given by $\delta\psi = \lambda^3_{\,\, a'} \, d\beta^{a'}$, where $(\beta^1, \beta^2) = (\alpha, l)$. The function $\lambda^3_{\,\, a'}$ is explicitly defined in (\ref{lambdaiaexplicit}). Since there is no force, it becomes simply
\begin{equation}
\lambda^3_{\,\, a'} = \frac{1}{M_0} \frac{\partial Q^{30}}{\partial \dot\beta^{a'}} = \frac{4}{3M_0r^2} \frac{\partial J^{2320}}{\partial \dot\beta^{a'}} \,\, ,
\end{equation}
where $M_0 = m_0 + 2m_1$. Using expression (\ref{bipodJ2320}) for $J^{2320}$ in terms of $l$ and $\alpha$, we get
\begin{align}
& \lambda^3_{\,\, \alpha} = - \frac{4}{3r^3} \frac{m_0 m_1}{(m_0 + 2m_1)^2} l^3 \si\alpha \\
& \lambda^3_{\,\, l} \, =\,  0 \,\, ,
\end{align}
which gives precisely the $\delta\phi$ in (\ref{bipoddeltaphi}).

Finally, it interesting to note that this swimming behavior can only be observed when the bipod is released from rest (with respect to the time symmetry). When $u$ has an initial component in, say, direction $\psi$, the torque components become
\begin{align}
&Q^{10} = 0
\nonumber\\
&Q^{20} = u^k v_k\frac{4 u^0}{3r^2} \Big[ u^0 \bar J^{0323} - u^3 \bar J^{0203} \Big]
\nonumber\\
&Q^{30} = u^k v_k\frac{4}{3r^2} \Big[ \left[ (u^0)^2 + (u^3)^2 \right] \bar J^{0232} + \Big.
\nonumber\\
&\Big. \qquad\qquad + u^0 u^3 \left( \bar J^{0202} + \bar J^{2323} \right) \Big] \, .
\end{align}
Since there is no torque propulsion in direction $e_1$, the body still does not leave the great sphere, as expected. Particularizing to the case of the symmetric bipod, the torque component $Q^{20}$ vanishes, showing that the bipod only moves along its symmetry axis (i.e., in $\psi$ direction). Clearly these torque components reduce to those in (\ref{bipodQi0}) when $u = e_0$. If $u$ has a non-negligible $u^3$ projection, then we see that the torque component $Q^{30}$ contains a tension contribution coming form $\bar J^{2323}$. Thus, if a bipod is set to move relativistically fast on a frictionless 2-sphere, it will not swim at low Reynolds number, unlike what happens in the Newtonian regime.

\section{A tripod in FLRW spacetime}
\label{sec:FLRW}

As we have seen, swimming at low Reynolds number in the curved spacetimes of general relativity is a non-general phenomenon, only expected to happen in very particular situations. Highly symmetrical spaces, nonetheless, should be good candidates to allow swimming, at least for certain body configurations and settings.  
In this appendix we consider spatially homogeneous isotropic spacetimes, called Friedmann-Lema\^itre-Robertson-Walker (or FLRW), showing that swimming is possible.

The FLRW metric has the form
\begin{equation}
g = d\tau^2 - a(\tau)^2 \left\{ \begin{array}{l}
d\psi^2 + \sin^2\! \psi \left( d\theta^2 + \sin^2\! \theta \, d\phi \right) \\
dx^2 + dy^2 + dz^2 \\
d\psi^2 + \sinh^2\! \psi \left( d\theta^2 + \sinh^2\! \theta \, d\phi \right)
\end{array} \right. \,\, ,
\end{equation}
where $\tau$ is the time as measured by the {\sl isotropic observers} (who follow the congruence of curves defining the isotropy of the space), and the coordinates have been respectively chosen as the spherical coordinates, Cartesian coordinates and hyperbolic coordinates.
The only independent curvature tensor components, in an orthonormal basis aligned with isotropic observers, are given by
\begin{align}
& R_{0101} = R_{0202} = R_{0303} =: \rho_0 \\
& R_{1212} = R_{2323} = R_{3131} =: \rho_1 \,\, ,
\end{align}
where $\rho_0$ and $\rho_1$ are functions of $a(\tau)$, and thus functions of $\tau$. 

For each Killing field $\xi$, equation (\ref{chi}) gives
\begin{equation}
\xi_k \hat F^k+ \frac{1}{2} \nabla_k \xi_l Q^{kl} = 0 \,\, ,
\end{equation}
where $\hat F^k$ is the non-spin-generated force, 
$\hat F^k := F^k - \frac{1}{2} v^{l} S^{qr} {R^k}_{lqr}$. Hence, the geometrical symmetries impose a constraint on the contributions of all the multipole moments (of quadrupole order or above) to the force and torque. Namely,
\begin{align}
&\hat F^1 = \hat F^2 = \hat F^3 = 0 \\
&Q^{12} = Q^{23} = Q^{31} = 0 \,\, .
\end{align}
Let us consider the case in which the body begins at rest with respect to the isotropic observes,
\begin{equation}
u(s_0) = e_0 \,\, ,
\end{equation}
and spinning in an arbitrary direction. The equation of motion (\ref{dotzapp2A}) for small integration times, in normal coordinates, then gives
\begin{equation}
\frac{d}{ds} \left[ \dot z^i - \frac{1}{M_0}  Q^{i0} \right] =  - \Gamma^i_{\,\, 00}(z) \,\, ,
\end{equation}
At $(l/r)^2$ order, we have
\begin{align}
Q^{10} &= \frac{4}{3}(\rho_0 + \rho_1) \left( \bar J^{0212} + \bar J^{0313} \right) \\ 
Q^{20} &= \frac{4}{3}(\rho_0 + \rho_1) \left( \bar J^{0121} + \bar J^{0323} \right) \\ 
Q^{30} &= \frac{4}{3}(\rho_0 + \rho_1) \left( \bar J^{0131} + \bar J^{0232} \right) \,\, ,
\end{align}
showing that there are only momentum-related quadrupole contributions to the torque. Thus, it must be possible to swim at this order.

Let us particularize to the case of the tripod described in \cite{Rodrigo}. If its axis is aligned with direction $e_1$ (without loss of generality, as the space is isotropic), then its internal symmetries simplify the torque components to
\begin{align}
Q^{10} &= \frac{8}{3}(\rho_0 + \rho_1) \bar J^{0212}  \\ 
Q^{20} &= 0 \\ 
Q^{30} &= 0  \,\, .
\end{align}
From expression (\ref{Jijkl}), this quadrupole moment is given by
\begin{equation}
\bar J^{0212} = - \frac{9}{8} m_1 l^3 \kappa_0 \si\alpha \,  \dot\alpha \,\, .
\end{equation}
where $l$ is the length of the arms and $\alpha$ is the angle between each arm and the axis of the tripod; also, $\kappa_0 = m_0/(m_0 + 3m_1)$, where $m_0$ is the mass at the pivot and $m_1$ are the masses at the extremities.
The differential additional displacement $\delta z^1$ then reads
\begin{equation}
\delta z^1 = - (\rho_0 + \rho_1) \kappa \kappa_0 l^3 \si\alpha \, d\alpha \,\, ,
\end{equation}
Finally, the net additional displacement for the cycle $(l, \alpha) \rightarrow (l + \Delta l, \alpha) \rightarrow (l + \Delta l, \alpha+ \Delta \alpha) \rightarrow (l , \alpha+ \Delta \alpha) \rightarrow (l , \alpha)$ is
\begin{equation}
\Delta z^1_{cycle} = -3 (\rho_0 + \rho_1) l^2 \kappa \kappa_0  \si\alpha \, \Delta l \Delta\alpha \,\, ,
\end{equation}
showing that it is indeed possible to swim at ``low Reynolds number'', at order $(l/r)^2$, in FLRW spacetimes.

It is interesting to notice that these spacetimes can become maximally-symmetric for particular choices of the function $a(\tau)$. For instance, one can choose $a(\tau) = r \sinh (t/r)$ in the hyperbolic case to obtain a {\sl de Sitter} spacetime. Then, $\rho_0 = - \rho_1$, so that the differential additional displacements vanish. This is compatible with the fact that there is no swimming in maximally-symmetric spaces.

\bibliography{swimming}

\end{document}